\pdfoutput=1
\documentclass[%
 rsi,
 amsmath,amssymb,
preprint,%
]{revtex4-1}

\usepackage{graphicx}% Include figure files
\usepackage{dcolumn}% Align table columns on decimal point
\usepackage{bm}% bold math
\usepackage[utf8]{inputenc}
\usepackage[T1]{fontenc}
\usepackage{mathptmx}

\begin{document}

%\preprint{AIP/123-QED}
\preprint{doi: 10.1063/5.0018926}

\title[A new setup for laboratory astrophysics]{A new multi-beam apparatus for the study of surface chemistry routes\\ to formation of complex organic molecules in space}
% Force line breaks with \\

\author{E. Congiu}
\email{emanuele.congiu@observatoiredeparis.psl.eu}
% \altaffiliation[]{Physics Department, University of Cergy-Pontoise.}%Lines break automatically or can be forced with \\
\author{A. Sow}%
%\affiliation{ 
%Authors' institution and/oraddress
%\\This line break forced with \textbackslash\textbackslash
%}
\author{T. Nguyen}%
\altaffiliation[Present address: ]{Institute of Low Temperature Science, Hokkaido University, Sapporo, Hokkaido, Japan.}
%\affiliation{ ... }
%\\This line break forced with \textbackslash\textbackslash
%}
\author{S. Baouche}
% \altaffiliation[Also at ]{Physics Department, University of Cergy-Pontoise.}

\author{F. Dulieu}
% \homepage{http://www.Second.institution.edu/~Charlie.Author.}
\affiliation{
{CY Cergy Paris Universit\'e, Sorbonne Universit\'e, Observatoire de Paris, PSL University, CNRS, LERMA, F-95000 Cergy, France}
%Second institution and/or address%\\This line break forced% with \\
}%

% We are supposed to use a one line address... this one.

\date{on November 7, 2020 accepted for pubblication in Rev. Sci. Instrum.}% It is always \today, today,
             %  but any date may be explicitly specified

\begin{abstract}
A multi-beam ultra-high vacuum apparatus is presented. In this article we describe the design and construction of a new laboratory astrophysics experiment -- VErs de NoUvelles Synthèses (VENUS) -- that recreates the solid-state non-energetic formation conditions of complex organic molecules in dark clouds and circumstellar environments. The novel implementation of four operational differentially-pumped beam lines will be used to determine the feasibility and the rates for the various reactions that contribute to formation of molecules containing more than six atoms. Data are collected by means of Fourier transform infrared spectroscopy and quadrupole mass spectrometry. The gold-coated sample holder reaches temperatures between 7 and 400~K.
The apparatus was carefully calibrated and the acquisition system was developed to ensure that experimental parameters are recorded as accurately as possible. A great effort has been made to have the beam lines converge towards the sample. Experiments have been developed to check the beam alignment using reacting systems of neutral species (NH$_3$, H$_2$CO). Preliminary original results were obtained for the \{NO~+~H\} system, which shows that chemistry occurs only in the very first outer layer of the deposited species, that is the chemical layer and the physical layer coincide. % FD : The same result with H2CO+H
This article illustrates the characteristics, performance, and future potential of the new apparatus in view of the forthcoming launch of the James Webb Space Telescope. We show that VENUS will have a major impact through its contributions to surface science and astrochemistry.

\end{abstract}

\pacs{Valid PACS appear here}% PACS, the Physics and Astronomy
                             % Classification Scheme.
%Use showkeys class option if keyword
                              %display desired
\maketitle

\section{\label{sec:intro}Introduction}

%GENERAL INTRO:
The detection of nearly 200 different molecular species over the last 50 years demonstrates that the interstellar medium (ISM) is home to a rich chemistry~\cite{McGuire2018}. In star-forming regions a multitude of species has been found and some of them, the so-called prebiotic species, are thought to be involved in the processes leading to the early forms of life. 
As gas and dust gradually condense into colder and denser interstellar matter, chemical processes begin to produce molecules from the mainly atomic gas. Although formaldehyde-sized molecules (H$_2$CO) exist in diffuse and translucent clouds~\cite{Liszt2006}, the production of complex species seems to occur in the densest regions, where UV fields are considerably reduced and temperatures are below 15~K~\cite{Caselli2012}.
%FD{\bf where the UV flux sufficiently reduced}.

Models of steady-state or time-dependent gas-phase astrochemistry do not always manage to simulate the observed molecular abundances because physical conditions of interstellar regions can themselves be time dependent, and not all chemical processes are known. Therefore, surface reactions on sub-micrometer-sized dust grains are necessarily invoked for the formation of many organic molecules~\cite{Cazaux2016,Taquet2012}.

The chemistry on the surface of cold grains occurs mostly at low temperature ($\sim$10~K). In the innermost part of the clouds new molecules form in and on H$_{2}$O-dominated dust grain mantles, mainly via reactions of atomic hydrogen and atomic oxygen with molecules and radicals. These processes have been invoked by astronomers for many years now~\citep{Tielens1982}, but it is during the last two decades that laboratory evidence has confirmed that the bulk of interstellar ices is formed in the solid phase through surface reactions (water, methanol, carbon dioxide, formaldehyde, ammonia, formic acid)~\cite{Dulieu2010,Jing2011,Oba2010,Ioppolo2011,Noble2011,Minissale2013,Minissale2015,Hiraoka2002,WatanabeKouchi2002,Fuchs2009,minissale2014c,Ioppolo2014,Minissale2016b,Joshi2012,Zins2012,Pirim2014}. 

The solid-state formation of organic material is therefore of considerable interest, as efficient surface reaction
routes also provide a general recipe to form relatively complex (i.e., containing more than six atoms) organic molecules, referred to as COMs, in star- and planet-forming regions~\citep{Charnley2001,Caselli2012,Vasyunin2017}.

One of the most important categories of data that astronomers need from laboratory experiments is to determine the rates for the reactions that contribute to formation and destruction of molecules under interstellar conditions. Two main processes are evoked to explain the chemical diversity observed in space: reactions in the gas phase and reactions in solid state (i.e., at the surface of dust grains).
See Smith (2011)~\cite{Smith2011} for a review on laboratory experiments for gas-phase processes.

For solid-state laboratory astrochemistry, surface science techniques at ultra-high vacuum conditions are used to qualitatively and quantitatively study the chemical physical processes leading to new species. Temperature-controlled desorption, TPD~\cite{Collings2004MRAS}, and the novel temperature-controlled during exposure desorption, TP-DED~\cite{Minissale2016}, techniques are now routinely used to analyze thermal and non-thermal desorption of atoms and molecules, providing the number and nature of newborn species, as well as the binding energies that will be used in the models. On the other hand, Fourier Transform Infrared (FTIR) and UV spectroscopy techniques are used to analyze ice samples, pure and mixed, \textit{in situ}, which allows laboratory astrophysicists to trace the evolution of the sample from the deposition of radicals until the formation of new products.
Until recently, solid state experiments have been carried out using single or double atom/radical sources, hence with a limited number of reactants and restrictions in the design of experiments aimed at simulating the processes leading to COMs. A paper was published this year that describes a triple beam line experiment including a novel atomic carbon source, which extends the possibilities of studying C-atom chemistry and the formation of COMs as this is likely to occur in the low density phase of molecular clouds~\cite{Qasim2020}.

The problem of how to form COMs with high enough efficiency to account successfully for their abundance in the ISM is a major problem in modern astrochemistry and astrobiology. 
To date, many of the facts required to fully understand the formation of COMs are yet to be determined. How and under what conditions are these molecules produced?
How far does this chemical complexity can go?

This paper describes the design and performance of VENUS, a new apparatus conceived to study surface reactions and reaction-diffusion processes that lead to the formation of COMs via non-energetic processes under conditions approaching those of the ISM. The experiments take place under ultra-high vacuum (UHV)
conditions (base pressure $1\times10^{-10}$~mbar) in a stainless steel chamber. Through a multi-beam assembly, up to five impacting species can be directed onto the cold target, consisting of polycrystalline gold or of an analog of cosmic dust. The sample temperature can be controlled in the 7 -- 400~K interval; quadrupole mass spectrometry and FTIR spectroscopy are used to characterize the deposited reactants and the final products. In Section~\ref{sec:design} we describe in detail the various components of VENUS, in Section~\ref{sec:performance} we present the experimental performances of the apparatus, and finally in Section~\ref{sec:potential} we discuss the importance of these results with respect to astrochemistry in the solid phase.

%*****************
%INSTRUMENT DESIGN
%*****************
\section{\label{sec:design}Instrument design}

\subsection{\label{subsec:overview}Goals and overview}

This article describes the conception and performance of a new experimental facility constructed to investigate reactions between atoms and molecules at low temperatures occurring on substrates of astrochemical relevance.

\begin{figure}[b] %[scale=0.5] [width=0.5\hsize] 
	
	\includegraphics[scale=0.55]{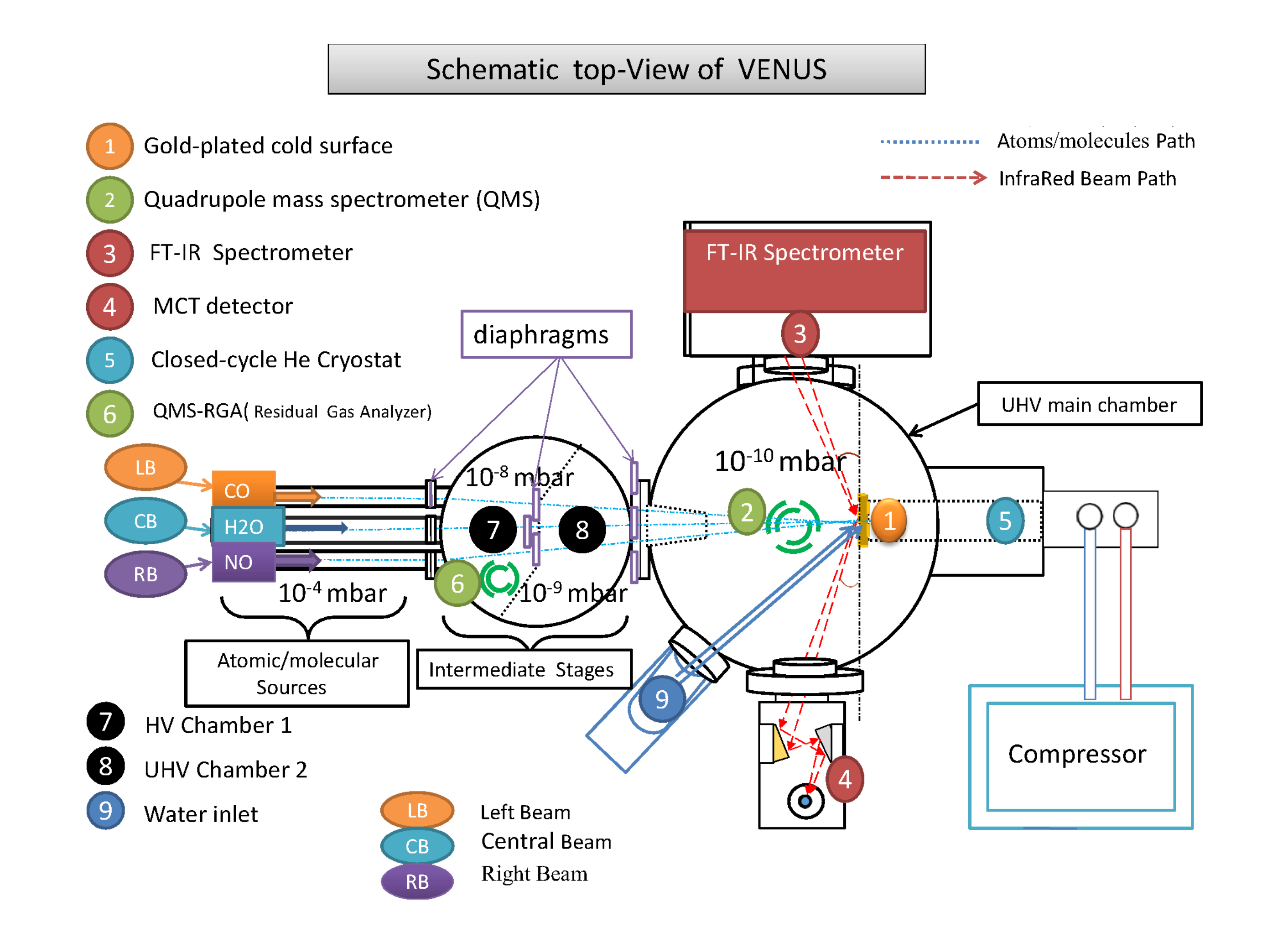}% Here is how to import EPS art
	\caption{A schematic diagram of the experimental setup.
	%[FD The figure is cut on the left side] 
	\label{fig:venus1}
	}
\end{figure}

The new setup called VENUS  -- an acronym of the French phrase \textit{VErs des NoUvelles Synthèses} (Towards new syntheses) -- was designed to serve as an upfront apparatus to study solid-state pathways leading to complex organic molecules in space. 
With few exceptions, complex molecules have been mainly detected in the gas phase, but there is strong evidence that they can be formed in ice mantles on interstellar grains~\citep{Herbst_vanDishoeck2009}. The suggested mechanisms leading to solid-state formation of COMs are:

\textit{i)} Hydrogenation (H addition;~\cite{Charnley1995ApJ}); 

\textit{ii)} Formation after H atom addition/abstraction on grains~\cite{Chuang2016MNRAS};

\textit{iii)} Radical-radical surface reactions (at T$>$30~K~\cite{Garrod_2008});

\textit{iv)} Cosmic-ray (CR) induced radical diffusion~\cite{Reboussin2014MNRAS};

\textit{v)} Impulsive spot heating on grains by CRs~\cite{Ivlev_2015ApJ},

The regions where the formation of COMs occurs generally have temperatures ranging between 10 and 100~K and typical densities of 10$^{4-5}$ cm$^{-3}$ (corresponding to pressures $\leq$~10$^{-12}$~mbar).

While modern technology allows us to easily reach very low temperatures, around 5~K or less if proper shielding is used, the technological limit for the lowest pressure is between 10$^{-11}$ and 10$^{-12}$ mbar, thus well above the "vacuum" conditions of the densest regions of the ISM. However, the operating conditions of ultra-high vacuum (UHV), defined as pressures below $10^{-9}$~mbar, are suitable for conducting experiments under conditions that simulate the solid-state synthesis of molecules in the ISM, because at these pressures the accretion time of typical pollutants (H$_{2}$, H$_{2}$O, CO) is much longer than the average duration of an experiment. At a pressure of 10$^{-10}$~mbar, it takes approximately two and a half hours for a monolayer of H$_{2}$ to form on the surface (assuming a surface density of sites $\sim 10^{15}$ cm$^{-2}$, a collision rate of $\sim 10^{11}$ particles cm$^{-2}$ s$^{-1}$~\cite{woodruffdelchar1994}. This allows time to perform experiments on a surface which is clean and well characterized, with only a few percent of a monolayer contamination per hour.
 
Therefore, to take into account the mechanisms of formation of COMs  that do not involve energetic processes (i.e., mechanisms \textit{i}, \textit{ii}, and \textit{iii}) and to approximate the conditions of the ISM in the laboratory, we had to employ ultra-high vacuum technology and the use of cryogenics. With this in mind, a new setup was designed and constructed for the study of sub-monolayer systems in which two or more (up to four) species could simultaneously react on a cold substrate of astrophysical interest.

VENUS consists of a multi-beam stage and three custom-built stainless steel chambers sealed with Conflat{\tiny$^{^{\circledR}}$} (CF) flanges (Figure~\ref{fig:venus1}). The multi-beam stage is made up of five independent small chambers serving as gas sources, the output of which enters the intermediary HV chamber 1 through a 2 mm diaphragm. The intermediary HV chamber 1 (CH1) and UHV chamber 2 (CH2) are built out of a single stainless steel cylinder and are separated by a S-shaped interior wall. Chambers CH1 and CH1 serve as a compact two-stage differentially-pumped path for the beam lines.  The third and final stage is a UHV chamber with a base pressure of 2 $\times$ 10$^{-10}$ mbar, that is the "reaction" chamber and we will refer to it as the \textit{main chamber} henceforth.

The main chamber contains a quadrupole mass spectrometer (QMS) and a gold-coated copper sample holder attached to the cold finger of closed-cycle He cryostat. The gold-coated target temperature can be varied between 7~and 400~K via a computer-controlled resistive heater. 
Although the minimum temperature of the surface is 7~K, this is sufficient to model interstellar chemistry, as molecular clouds are estimated to have temperatures in the region of 10~K~\cite{Williams_Herbst2002SurSc}.

At present, VENUS has four operational triply differentially-pumped beam lines having fluxes of about $2 \times 10^{12}$ molecules~cm$^{-2}$~s$^{-1}$, which means that one monolayer (1 ML, corresponding to $10^{15}$ molecules~cm$^{-2}$) is grown on the gold-coated mirror in about 10 minutes. Deposited and newly formed species are monitored \textit{in situ} by means of Fourier transform reflection absorption infrared spectroscopy (FT-RAIRS) using vibrational fingerprint spectra in the mid-infrared (2.5~--~15~$\mu$m) domain.

The VENUS setup has been optimized for the study of very thin layers (sub-monolayers to one monolayer) of adsorbates, although systems of thick ices (pure and mixed) can be investigated as well. 
%-- without energetic sources other than thermal and chemical energy (mechanisms i), ii) and iii) --
The combination of all these characteristics allows VENUS to perform controlled deposition of very thin layers and to monitor the reaction kinetics in real time, making this experimental facility an ideal tool for studying sub-monolayer systems of several reactants and the formation of complex molecules in the solid phase.

%************************

\subsection{\label{subsec:mainch}Intermediary stages and the main chamber}

\begin{figure}[b] 
%[scale=0.5] [width=0.5\hsize] 
\includegraphics[scale=0.2]{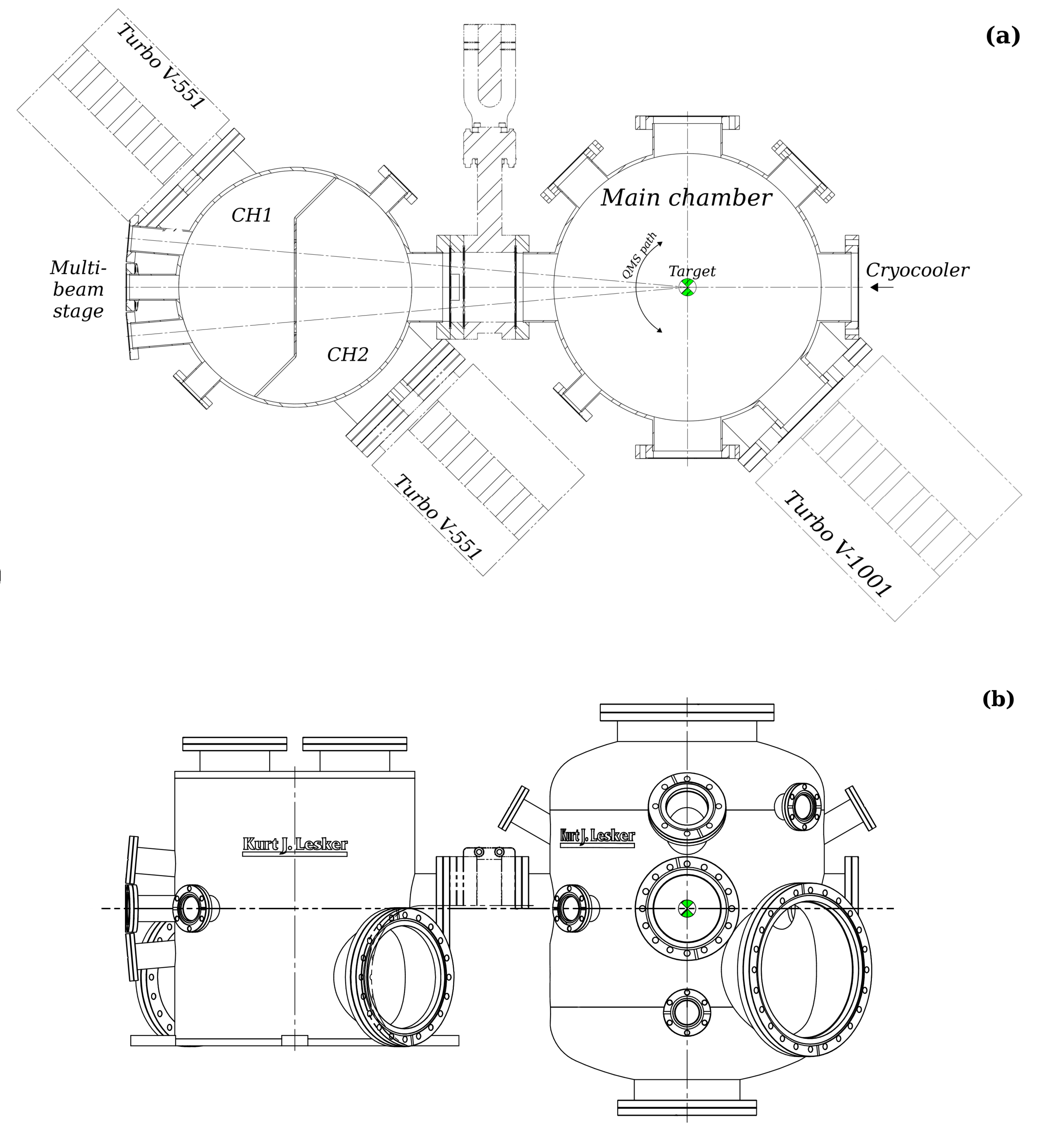}
\caption{Schematic top-view (a) and side-view (b) diagrams of the vacuum chambers showing the position of various experimental components. \label{fig:chambers}
	}
\end{figure}

The apparatus consists of three vacuum chambers as shown in Figure~\ref{fig:chambers}. The first two chambers, CH1 and CH2, are intermediary chambers that serve as two independent stages of differential pumping for the beam lines. CH1 and CH2 are separated by an S-shaped stainless shield (3 mm in thickness) and each chamber is pumped by a turbo-molecular pump (Agilent Technologies, Turbo-V 551 Navigator) having nominal pumping speed of 550~l~s$^{-1}$.
CH1 has a base pressure of $10^{-8}$~mbar monitored by a full-range pirani/cold cathode gauge (Pfeiffer Vacuum, PKR251), and contains a Residual Gas Analizer that is used for the acquisition of mass spectra of the incoming species injected via the beam line ports at left. The intermediary chamber CH2 has a base pressure of $2 \times 10^{-10}$ mbar measured with a cold catode gauge (Pfeiffer Vacuum, IKR270). 

A pneumatic gate valve separates the intermediary stages from the UHV chamber via the frontal CF 100 flange. The main chamber, illustrated in Figure~\ref{fig:venus1} and~\ref{fig:chambers}, is a custom-made non-magnetic stainless steel cylinder 60 cm tall, with a diameter of 50 cm, built by {\footnotesize{\copyright}}Kurt J. Lesker Company. A turbo-molecular pump with pumping speed 1000 l s$^{-1}$ (Agilent Technologies, Turbo-V 1001 Navigator) maintains UHV conditions in the chamber, at a base pressure of a few $10^{-10}$ mbar, if an appropriate baking of the chamber is performed at $\sim$70~C$^\circ$ for about 3 days. The Turbo-V 1001 pump is backed via a primary vacuum pipe (<$10^{-3}$ mbar) that runs above the apparatus and is evacuated by a roots pump (Pfeiffer Vacuum, Okta 500), located in the basement, with a nominal pumping speed of 280 to 840~m$^{3}$~h$^{-1}$. 
%Ti sublimation pump
The main chamber is additionally pumped with a three-filament titanium sublimation pump (MECA 2000, PFT.3), which is switched on to achieve optimum UHV conditions whenever the hydrogen load is high.\\
A cold catode gauge (Pfeiffer Vacuum, IKR270) is used to monitor the main chamber pressure.

\begin{figure}[b] 
	%[scale=0.5] [width=0.5\hsize] 
\includegraphics[scale=0.3]{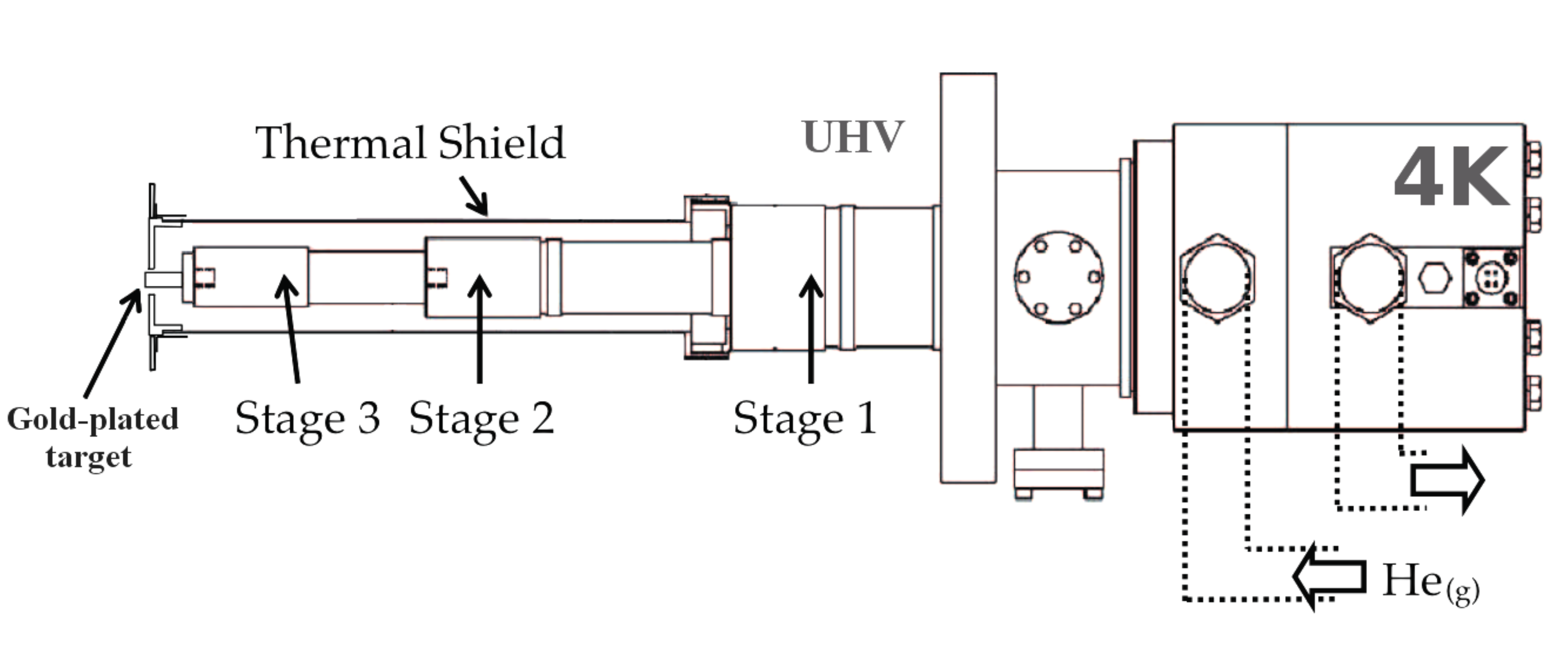}
\caption{Detailed schematic of the cryocooler and the target surface. \label{fig:cryocooler}
	}
\end{figure}

At the rear CF 100 port of the main chamber is mounted the cryocooler, shown in Figure~\ref{fig:cryocooler}. The sample holder is an OFHC (Oxygen-Free High Conductivity) copper cylinder (9 mm in diameter) mounted at the end of the cold finger of a helium-cooled cryostat (Advanced Research
Systems 4K DE-210SB cryo-cooler coupled to an ARS-10HW
compressor, with a typical cooling power of 0.4~W
at 10~K). The sample holder is mounted on
the third stage of the cryostat, and this whole region is thermally isolated
from the environment by a thermal shield. The protective thermal cover consisting of a nickel-coated copper
(internal) and stainless steel (external surface) cylinder surrounding the whole assemblage is attached to the first stage of the cryostat ($\sim$50~K).

Three different sensors are used to measure the temperatures in different parts of the cryocooler: a KP-type thermocouple (Au 0.07\%-Fe/Chromel); a K-type
thermocouple (Nickel-Chromium/Nickel-Aluminium); and a silicone diode (DT 470). The three sensors are connected on the thermal shield, on the second stage of the cryostat, and on the sample holder, respectively. A heating resistor too is attached to the back of the OFHC Cu sample holder. All temperature sensors and the resistor are connected to a Lakeshore 340 temperature controller.

The gold-plated target, that is the sample holder 9~mm circular base facing the beam lines, is optically flat, chemically inert, and ensures optimal mid-infrared light reflectance. The surface temperature is measured with the DT~470 silicon diode and is regulated by the Lakeshore 340 controller to $\pm$0.1 K
with an accuracy of $\pm$2~K in the 8 -- 400~K range, as determined by calibration carried out monitoring the multilayer desorption temperature of CO and N$_2$~\citep{Thanh2018AA}.

The Lakeshore 340 is controlled remotely by a homemade Labview program which allows us to perform the following procedures:

1) regulating heating ramps and final temperature during cooling or heating of the sample;

2) monitoring and recording the temperatures captured by the sensors during an experiment;

3) adjusting the heating power in real time to ensure a linear increase in surface temperature.

The QMS (Hiden 51/3F) is mounted in an upright position to a rotatable CF 40 port soldered off-axis to a differentially pumped rotary CF 100 flange, attached to the bottom port of the main chamber. This configuration allows the mass spectrometer to be rotated around its axis and also to move along a semicircular path around the target (see Fig.~\ref{fig:chambers}). Also, the QMS can be translated vertically, which allows us to align the QMS and the sample with any of the beam lines. The QMS is used for three main purposes: in its low position, to analyse
residual gas in the main chamber as a function of atomic or molecular mass (in the range 1 -- 300~amu); in its high position, facing any of the beam lines, to characterize the chemical nature of the beam as well as the beam intensity, and thus to check the accuracy of the beam alignment;
and again in its high position, facing the target, the mass spectrometer is used to measure the desorption of species from the surface during a TPD. The QMS is controlled via a software program provided by Hiden. Technical settings, such as the energy of the ionizing electrons and the dwell time for each mass, can be controlled via this interface. It also records both the masses being measured and the temperature of the surface, input from the Lakeshore 340 as an external signal on the auxiliary entry.

The target surface is located approximately in the center of the main chamber, at the point where both the four beam lines and the infrared beam converge. The whole cryocooler is mounted on a 3-axis translational stage,  which allows translations and fine adjustments of the surface position both in the z (in and out of the main chamber) and the x, y (horizontally and vertically, respectively, relative to the QMS and beam lines) directions.

\subsubsection{\label{subsubsec:ices}Growth of water ice substrates that mimic interstellar icy mantles}

Amorphous solid water (ASW) is reputed to be the most
abundant form of water in the Universe, thanks to its
propensity for forming, molecule after molecule, as a deposit
on interstellar dust particles.
In the laboratory, water ice film are obtained by the condensation of water vapour onto the cold target. The properties of the ice depend upon the incident angle of water molecules, the temperature of the surface both during deposition and
any increase in temperature experienced after deposition~\cite{Kimmel2001JChPh}. 
We constructed a water vapour delivery manifold for growing water ice onto the sample surface with control over the ice thickness and deposition rate. One UHV leak valve is mounted to one of the main chamber ports and is connected to a stainless steel gas-handling manifold ending with a 4-mm nozzle. 
The nozzle outlet can be positioned at a "high" position to flood the main chamber with water vapour without exposing the sample to the direct flow of molecules emerging directly from the nozzle. 
The ice films are grown by admitting water vapor into the main chamber so it freezes out on the gold target. This method, also known as "background deposition",
allows us to control the amount (and thickness) of the deposited ice film. The
water vapor is obtained from deionized water which has been
purified by several freeze-pump-thaw cycles, carried out under vacuum.
The rate at which molecules of water impinge upon the sample will be proportional to the partial pressure of water in the main chamber. By monitoring the partial pressure of the "flooding" species, ice thickness and deposition rate are computer controlled within the accuracy of the pressure gauge reading (i.e., 0.1 monolayers) using a custom-made LabView program.\\
Typically, water vapour is injected into the main chamber until a pressure of 1.2$\times$\(10^{-8}\) mbar is reached, and is kept constant during the deposition process until the desired ice thickness is attained. The purity of the water vapor, that is absence of air contamination, is monitored via the QMS.
The calibration of the water ice thickness was carried out via FT-RAIRS and QMS in a dedicated set of experiments where the amount of water ice deposited was measured as a function of time using the H$_2$O partial pressure in the chamber, since a given pressure corresponds to a measurable flux of molecules impinging on the walls of the chamber. Subsequently, the comparison between the number of deposited water layers (calculated by the LabView program) and the infrared/mass spectra allows definition of the one-monolayer standard expressed in IR-band area units or TPD peak area units, with an accuracy of  $\pm 0.2$ monolayers.\\

Three major ice morphologies are used to simulate the icy mantles covering the grains of dust in dark clouds:

-\textit{porous ASW}: mimics a very disordered substrate with a high effective surface area, because of the presence of cracks and pores having themselves an internal surface area. To produce porous ASW, water vapor is dosed while the surface is held at a constant temperature between 10 and 40~K, depending on the required degree of surface disorder.

-\textit{non-porous ASW}: mimics the compact ASW ice which comprises the bulk of
interstellar ice.
To produce non-porous ASW, water is dosed while the surface is held at a constant temperature of 110~K.

-\textit{crystalline water ice}: mimics the crystalline ice seen in some
star-forming regions. 
To form crystalline water films,
the surface is held at 120~K during the deposition, then flash heated at 50~K min-1 to 140~K, and finally at 10~K min-1 to 142.5~K. Complete crystallisation of the ice film is checked through verification of the infrared water band profile between 3400 and 3100~cm$^{-1}$, and thus cooled down to 130 K.\\

Once the deposition phase is complete, typically we wait between 30 and 60 minutes (depending on the thickness of the deposited ice) until the pressure in the main chamber has reached its pre-dosing value ($\sim 10^{-10}$~mbar), and only then the sample temperature is cooled further (typically to 10~K) to perform the subsequent experiments.\\

By lowering the water leak manifold to a "low" position, the nozzle is located 1~cm in front of the target, aims at the cold surface and the flow of water vapour can be directed at normal incidence onto the sample. Nevertheless, repeated testing of this configuration proved it was not suited for growing a uniform well characterized layer of water ice. It can however be used to quickly form a thick sublayer of compact ice ($>$100~ML) at a surface temperature of 110~K, on which the subsequent well characterized ice films are grown.

\subsection{\label{subsec:rairs}FT-IR spectrometer}

Composition, thickness, and morphology of the deposited ices and of the newly formed species may be monitored \textit{in situ} through absorption reflection spectroscopy (RAIRS) using a Bruker Vertex 70v spectrometer.

RAIRS was first applied many years
ago, when Pickering and Eckstrom (1959)~\cite{pic59} recorded spectra from carbon monoxide and
hydrogen adsorbed on metal films, using a multiple reflection arrangement in which the infrared beam was at near normal incident to the substrate.
Following the principles laid down by Greenler (1966)~\cite{gre66}, higher sensitivity can be obtained in a reflection experiment if grazing angles of incidence are employed~\cite{hollins94, trenary00, fraser02, bennett04}. FT-RAIRS turns out to be a valuable technique to conduct studies on molecules adsorbed on a surface, and provides information that cannot be inferred from TPD mass spectra. In fact, the combination of TPD and RAIRS provides a better understanding of the system than either technique alone. Besides, IR spectroscopy will allow us to investigate surface-formation processes concerning complex molecules, starting from the recently observed COMs such as methyl formate (HCOOCH$_3$) and dimethyl ether (CH$_3$OCH$_3$) to refractory carbon and nitrogen bearing molecules of astrophysical
and astrobiological importance, that cannot be investigated thoroughly through the TPD technique alone.

\begin{figure}[b] 
	%[scale=0.5] [width=0.5\hsize] 
	
	\includegraphics[scale=0.41]{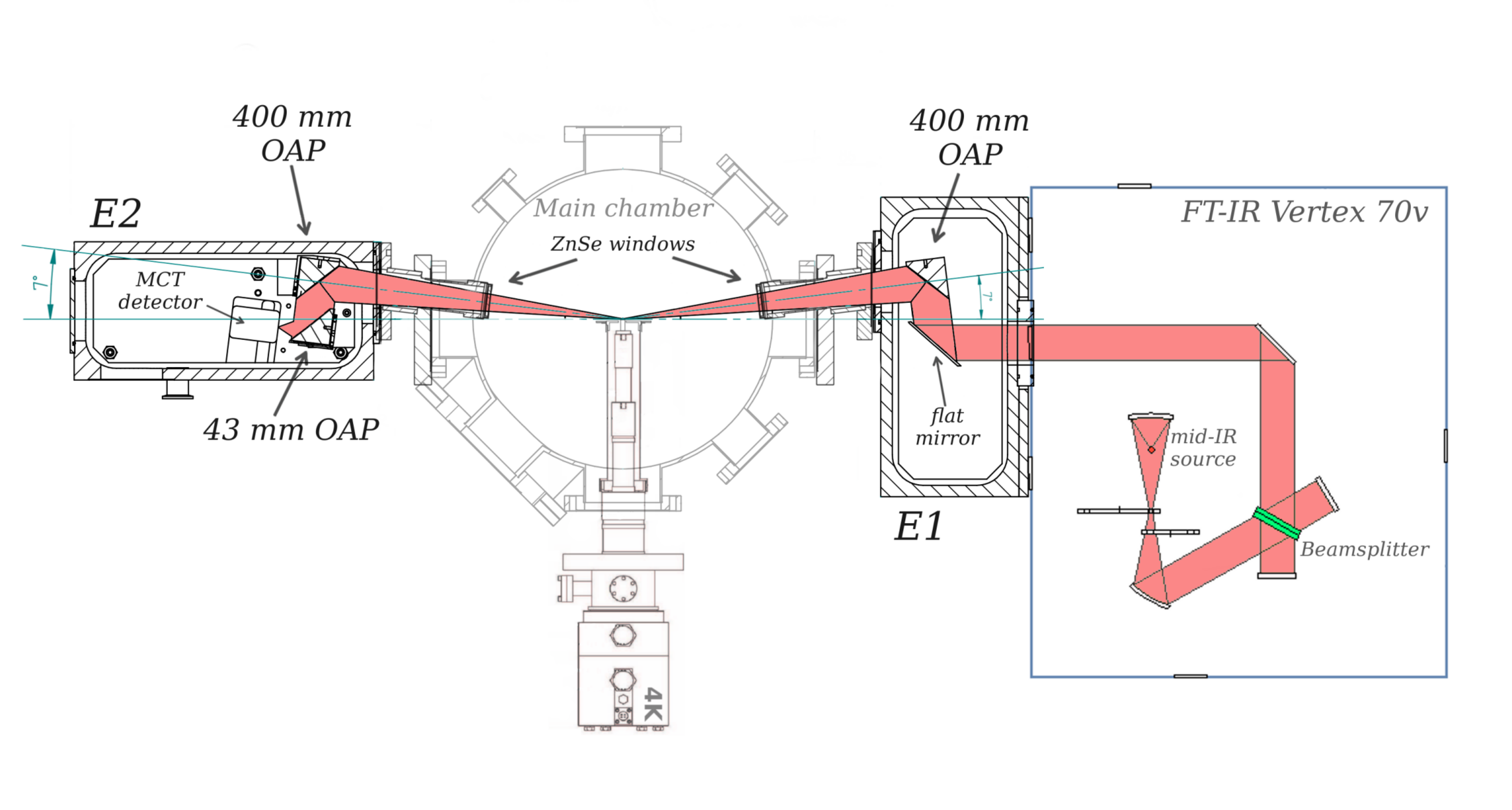}
	\caption{Detailed schematic of the FT-IR setup showing the infrared beam path and the position of various components located inside the pumped enclosures E1 and E2.  \label{fig:rairs}
	}
\end{figure}

The Vertex 70v Fourier transform infrared (FT-IR) spectrometer mounted on VENUS possesses airtight lids and windows so it can be kept under vacuum to avoid atmospheric H$_{2}$O and CO$_{2}$ spectral contamination. The FT-IR instrument is set up in a reflection-absorption infrared spectroscopy configuration and the spectra are recorded at a grazing incident angle of 83~$\pm 1^{\circ}$ (Figure~\ref{fig:rairs}). The parallel and unpolarized infrared beam (40~mm in diameter) exiting the FTIR is focused onto the sample using two gold-coated mirrors: a flat and a 90$^{\circ}$ off-axis parabolic (OAP) mirror (effective FL = 400~mm), mounted in a pumped enclosure (E1) adjacent to the main chamber and isolated from it with a ZnSe window. External to the UHV chamber, through an opposite ZnSe window, the infrared beam enters a second pumped enclosure (E2) after being reflected off the sample substrate coated in thin film of poly-crystalline gold that acts as a flat mirror. Inside E2, another 400~mm OAP mirror is used to collect the diverging infrared light coming from the sample, that is finally focused onto the sensor element of a liquid-N$_2$-cooled mercury cadmium telluride (MCT, spectral sensitivity 6000 -- 600~cm$^{-1}$, or 1.6 -- 16~$\mu$m) with a 43~mm OAP. Although copper and silver are also two highly reflective metal surfaces
used commonly for applications in the mid-IR spectral range, with a reflectance percentage comparable to that of gold, only gold-coated mirrors were used on VENUS because they are very broadband, from 0.7 to 10~$\mu$m they reflect more than 98\% of the incident radiation, and this percentage stays in this range up to 25 $\mu$m.

\begin{figure}[b] 
	%[scale=0.5] [width=0.5\hsize] 
	\includegraphics[scale=0.6]{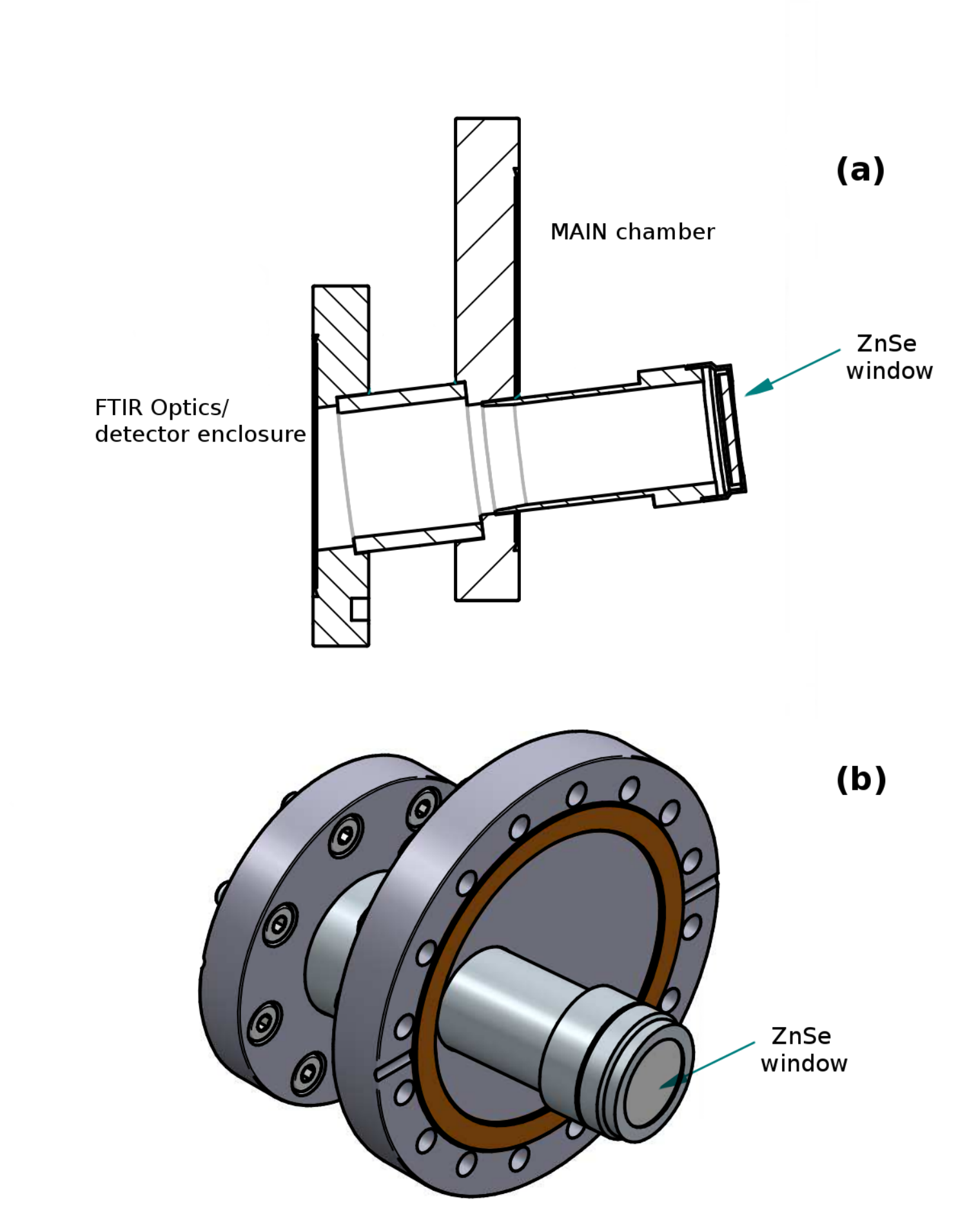}
	\caption{Schematic (a) and 3-D view (b) of the custom-designed double flange used to couple the FT-IR setup to the main chamber. \label{fig:RAIRSflanges}
	}
\end{figure}

To couple the external enclosures housing the optics and the MCT detector to the main chamber, two custom-designed flanges are employed, whose schematic is shown in Fig~\ref{fig:RAIRSflanges}. Each coupling piece, manufactured by Bruker, is made of two flanges soldered to a stainless steel tube that forms an angle of 7$^{\circ}$ with the flange normal. Such a design allows the infrared beam to enter the main chamber and be incident upon the sample with the desired grazing angle of 83$^{\circ}$. The spectrometer and the two external optical enclosures are pumped with a dry scroll pump (Edwards, model nXDS15i) that allows the pressure to be in the low 10$^{-2}$~mbar.

FT-IR spectra between 4000 and 650 cm$^{-1}$ (2.5 --~15 $\mu$m) may be acquired prior, during, or subsequently to either deposition or TPD experiment, with a typical resolution of 4 cm$^{-1}$. The number of averaged scans per saved infrared spectrum is typically 256, corresponding to an acquisition time of about 2 minutes. In case a shorter acquisition time is needed, such as real time monitoring of reaction kinetics during a TPD, the Vertex 70v spectrometer allows us to use a rapid scan mode at a reduced resolution of 8~and 16~cm$^{-1}$.

The FT-IR spectrometer is controlled via OPUS{\scriptsize$\mathrm{^{^{^{TM}}}}$}, a proprietary software program provided by Bruker. The spectral acquisition settings, such as the resolution, diaphragm aperture, number of scans, spectral range, performance tests, pumping and venting of the spectrometer, etc., can be controlled via this interface. The OPUS \textit{Measurement dialog} also allows us to start recording the background measurement (i.e., spectrum recorded without a sample being on the surface) and the sample spectrum. The result sample spectrum is automatically calculated by dividing the sample spectrum by the background measurement. The software is
used either to record a single spectrum or to start a repeated data collection, that is a series of automatic acquisitions recorded sequentially during a deposition, or during H atom bombardment of the sample, or during a TPD. During the experiments, the standard ceramic mid-infrared (MIR) source and a KBr beamsplitter (spectral range 4800 to 450~cm$^{-1}$) are permanently mounted on the spectrometer. However, a near-infrared (NIR) source and a quartz beamsplitter (15000 to 3300~cm$^{-1}$) are used whenever the system needs alignment of the optics (mirrors, target, and MCT sensor).

\subsection{\label{subsec:beams}Molecular and atomic beam lines}

There are currently four operational beam lines mounted on VENUS, although a fifth source could be implemented if necessary, according to future research programs.

The possibility of using four molecular/atomic beam lines is undoubtedly one of the most important properties of this apparatus, and a unique feature among astrochemistry laboratories throughout the world.

Each beam line is composed of three stages of differentially pumped chambers separated by diaphragms (whose positions are illustrated in Fig.~\ref{fig:venus1}), and was designed to deliver molecular and atomic species onto the sample at a rate low enough to maintain a pressure $< 10^{-9}$~mbar in the main chamber at all times.

\begin{figure}[b] 
	%[scale=0.5] [width=0.5\hsize] 
	\includegraphics[scale=0.8]{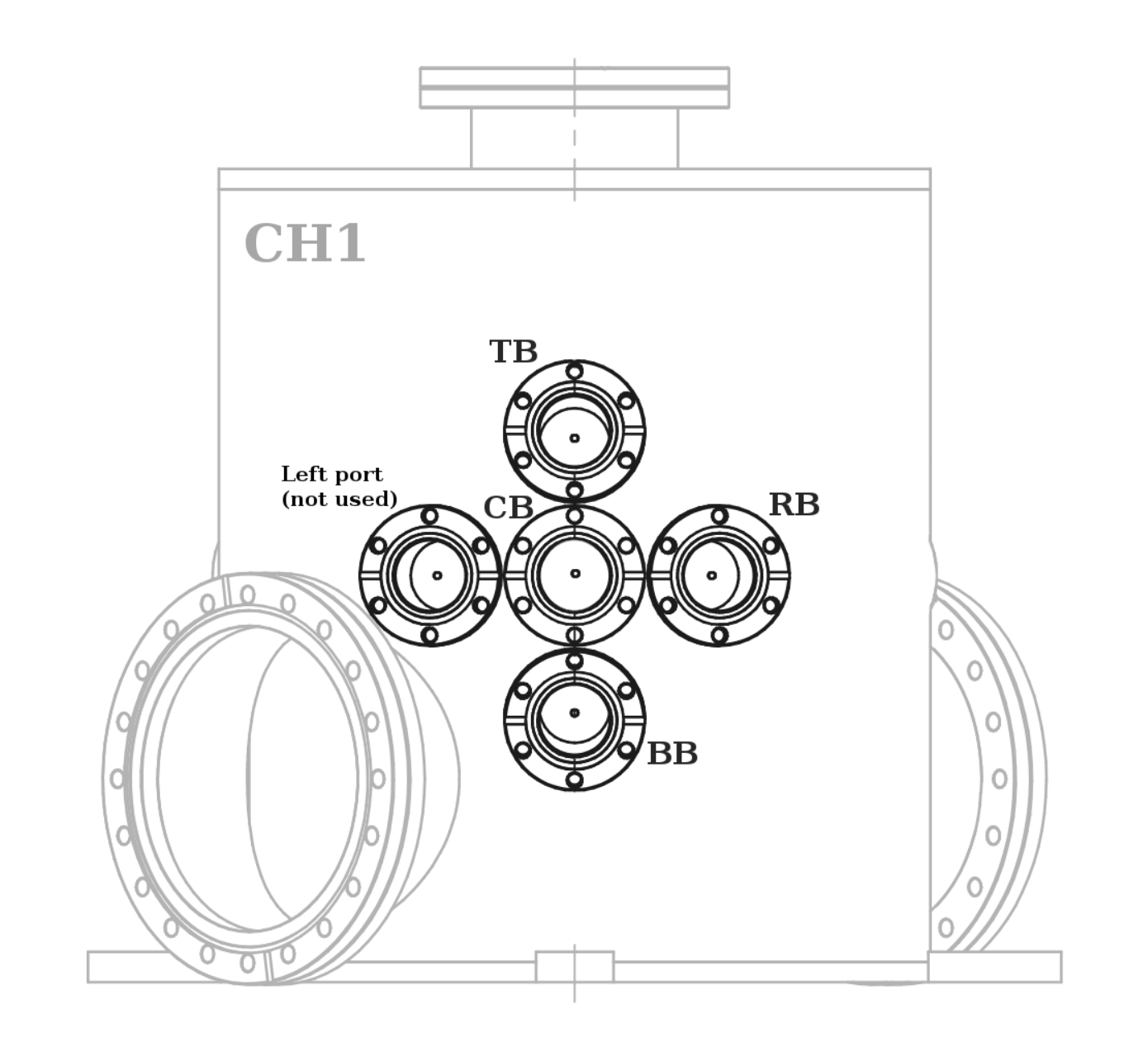}
	\caption{Frontal view of the apparatus showing the five ports to which the first stages the beam lines are attached. \label{fig:beamports}
	}
\end{figure}

The four first stages (or stage 1) of the beam lines compose the multi-beam stage. A frontal view of the apparatus, presented in Figure~\ref{fig:beamports}, shows the five CF 40 ports to which the beam line stage~1 chambers are attached. As mentioned above, only four ports are currently used, that is the central port (central beam, CB), the top port (top beam, TB), the right port (right beam, RB), and the bottom port (bottom beam, BB). The top, right, and bottom ports are soldered as to form an angle of 5$^\circ$ with the central port (normal to the upright axis) so that all the beam lines converge on an ideal point lying onto the central axis of the main chamber.

Stage~1 of each beam line is composed of a quartz tube (the source, 4~mm inner diameter), a vacuum chamber evacuated by a turbo molecular pump (Leybold Vacuum, Turbovac SL80, pumping speed 65~l s$^{-1}$), and a skimmer terminating with a diaphragm of 2 mm in diameter. The pressure of the beam in the quartz tube is typically $\sim$1~mbar, diminishing gradually to pressures of $10^{-4}$ -- $10^{-5}$~mbar within the chamber of the first stage to a value of $10^{-10}$~mbar in the main chamber. Collimators (diaphragms) of 3~mm diameter separate CH1 from CH2, and CH2 from the main chamber, so the beams are angularly resolved to a very small solid angle of $\sim 10^{-5}$ steradian.

\begin{figure}[b] 
	%[scale=0.5] [width=0.5\hsize] 
	\includegraphics[scale=0.71]{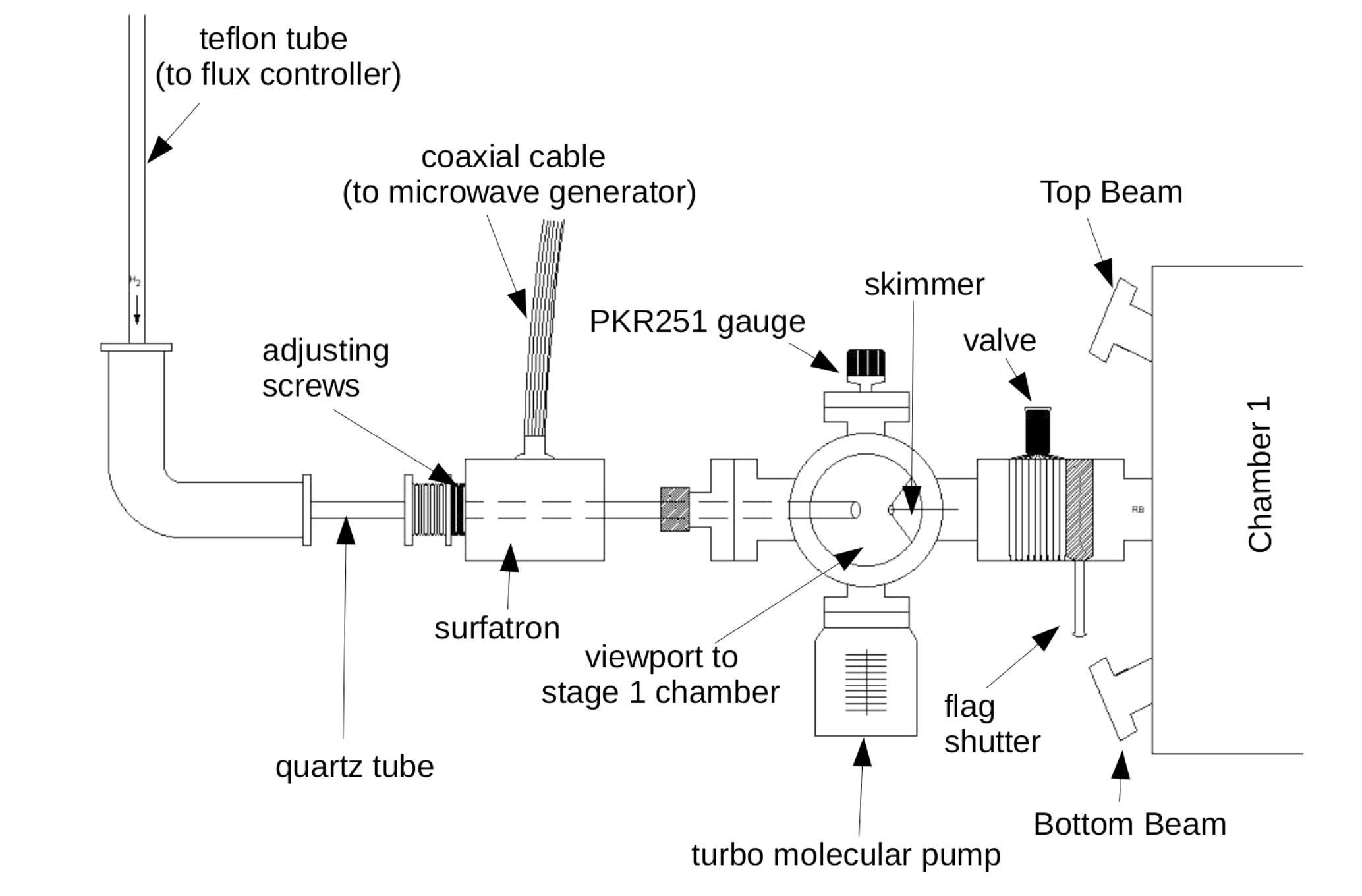}
	\caption{Schematic of stage 1 of the right beam (RB). Stage 1 of the other beam lines is essentially a replica of the RB first stage without the Surfatron and the flag shutter. \label{fig:RBstage1}
	}
\end{figure}

The right beam is also equipped with an air-cooled microwave cavity that surrounds the quartz tube (see Figure~\ref{fig:RBstage1}), and with a simple flag shutter. This flag shutter enables us to modulate the beam (let it through or block it) very quickly (within 1 second). 
A valve separating each of the four first stages from CH1 can be closed to prevent the penetration of the beam into the main chamber.

The flux of gaseous species entering TB, CB and BB sources is controlled by an automated regulator (Bronkhorst High-Tech control valve 1~sccm~\footnote{Standard cubic centimetre per minute: 1 sccm = 592\,m$^3$ Pa s$^{-1}$ in SI units.}) set at 10\% of maximal gas flow rate, while H$_{2}$ injected into the RB source is controlled with a 10~sccm regulator typically set at 15\% of its maximal flow rate.\\
The great advantage of the Bronkhorst flux controller over a manual needle valve is that a given molecular flow is accurately reproducible from one experiment to another, even for experiments conducted on different days.

Gas is introduced into the first vacuum chamber through the 4~mm inner diameter tube, producing an effusive beam, and passes through three chambers (stage 1, CH1, CH2) and into the main chamber via three diaphragms. It should be noted that an aligned beam line requires the source (the 25~cm long quartz tube), the three diaphragms, and the target to be perfectly aligned, which is checked beforehand by a laser beam to ensure that all beam lines are aimed at the same spot on the sample. In addition to maximizing the beam flux intensity, 
ideally it would be preferable that all the beams be perfectly overlapped on the sample, for a 100\% overlap allows maximum physical and chemical interaction between two or more different species deposited through the beam lines.

With the Bronkhorst control valve by-passed and replaced by a standard leak valve, the central beam is used to inject water vapor or any other substance contained in a glass vial that is liquid at room temperature and has a vapor pressure that can vary from a few mbars (i.e., H$_2$O, CH$_3$OH, HCOOH) to about 1 atm (i.e., CH$_3$CHO). The beams can be molecular or, if the microwave discharge is activated, the RB will be a mixture of atomic and molecular species. Due to recombination of atomic species on surfaces of the beam source, the dissociation efficiency, $\tau$, is never 100\%. $\tau$ is calculated via the formula:

\begin{equation}
\tau = \frac{S_{OFF} - S_{ON}}{S_{OFF}}
\end{equation}

where S is the number of ion counts of the parent molecular species measured with the QMS in front of the target when the discharge is switched off or on. The value of $\tau$ is calculated before every experiment involving atoms. Typically measured dissociation fractions are between 60~and 80\% for H$_{2}$ and D$_{2}$, $\sim$~50\% for O$_{2}$~\cite{Minissale2019AA}, and only $<$5\% for N$_{2}$, due to the greater strength of the nitrogen-nitrogen triple bond.

Air cooling proves to be essential, not only to reduce the discharge temperature, but also to inhibit recombination of hydrogen atoms on the walls of the discharge tube~\cite{wood62}. Microwave-frequency power from a SAIREM's GMS 200\,W 2450\,MHz microwave generator is transmitted to the gas via a Surfatron 2450\,MHz (SAIREM), generating a plasma upon application of a spark. Impedance matching of the microwave cavity is performed using the adjustment screws on the Surfatron to ensure the maximum transmission of forward power (typically, 60\,W are used to dissociate H$_2$/D$_2$) and to reduce the reflected power.

Molecules and atoms delivered from the beam line to the main chamber are at ambient temperature of $\sim$300~K, even if produced in a discharge, due to the transfer of kinetic energy by collisions in the source. Besides, in an effusive source, the molecules in the beam have temperatures close to that of the tube.. 
Also, we previously determined that atoms are in their ground state~\citep{Amiaud2007, Congiu2009}.

\subsection{\label{subsec:calibration}Calibration procedures}

\subsubsection{\label{subsubsec:beamflux}Beam flux}

One of the first tests we performed was to check the relative intensity of the four beams using carbon dioxide, a highly active mid-infrared species. 

\begin{figure}
	\scalebox{1}{{
			\includegraphics[scale=0.34]{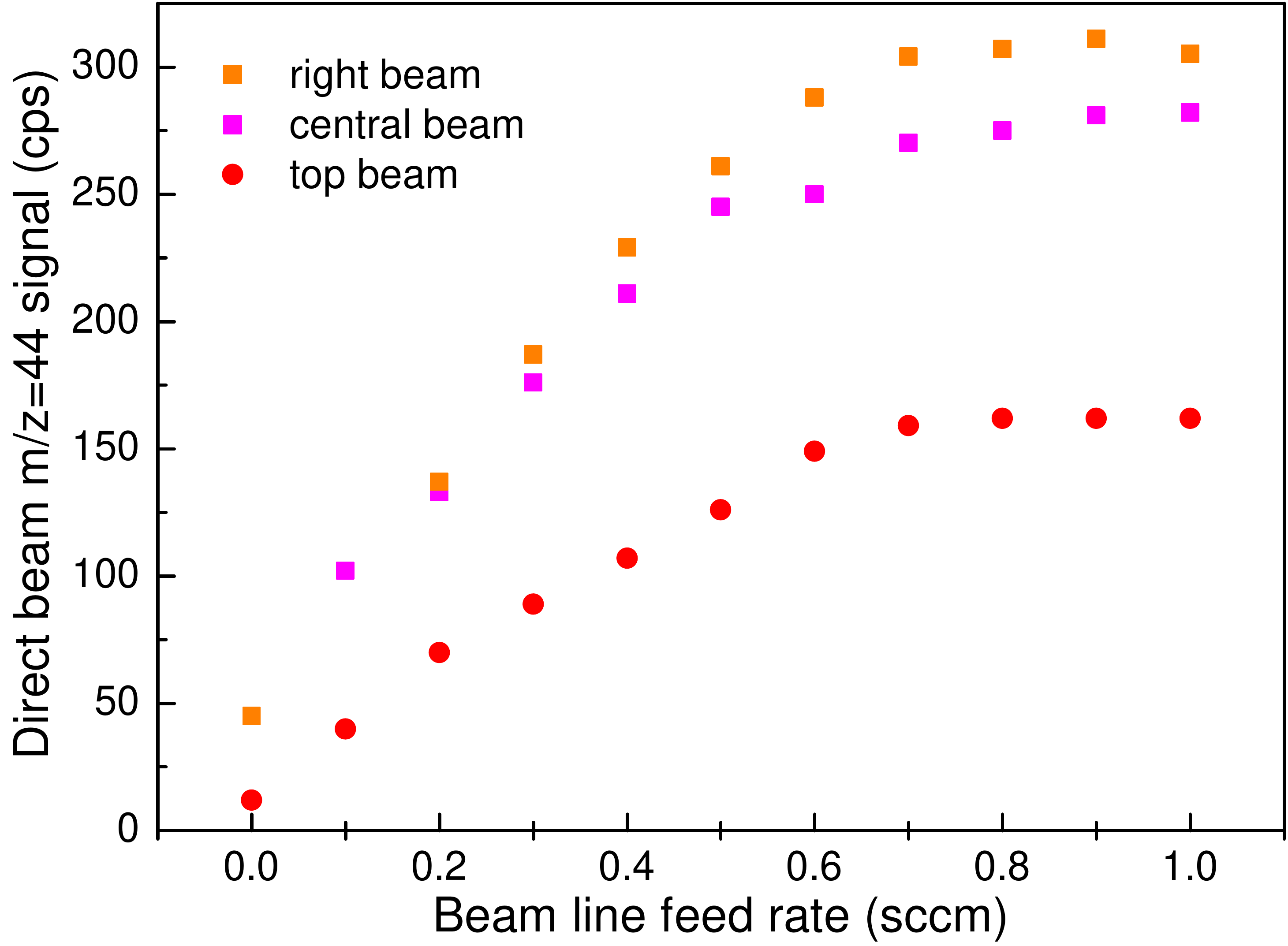}}}% Here is how to import EPS art
	\caption{\label{fig:beams1} Beam intensity as a function of CO$_{2}$ source feed rate in sccm for three of the operational beam sources.}
\end{figure}

Figure~\ref{fig:beams1} shows the intensity of the CO$_{2}$ beam flux expressed in ion counts s$^{-1}$ as a function of source gas supply in sccm, measured with the QMS facing each beam at a time.
The right and central beam have a comparable slope, namely these two beam lines have very similar flux intensities. The saturation value, that is the threshold value where proportionality between the source pressure and the beam line flux is lost, is reached at a source feeding rate of 0.5~sccm for the central beam, and 0.6~sccm for the right beam. The top beam appears to be less intense, which is likely due to a lesser precise alignment between the quartz tube and the diaphragms. The onset of the saturation region is generated by accumulation of molecules within the first stage chambers (see Figure~\ref{fig:RBstage1}), which makes the resulting mean free path of effusing molecules comparable or even shorter than the dimensions of the chamber itself. We thus determined that source feed rates $<0.5$~sccm must be used to avoid flooding of the first stage and blocking the beam path.

\begin{figure}
	\scalebox{1}{{
			\includegraphics[scale=0.31]{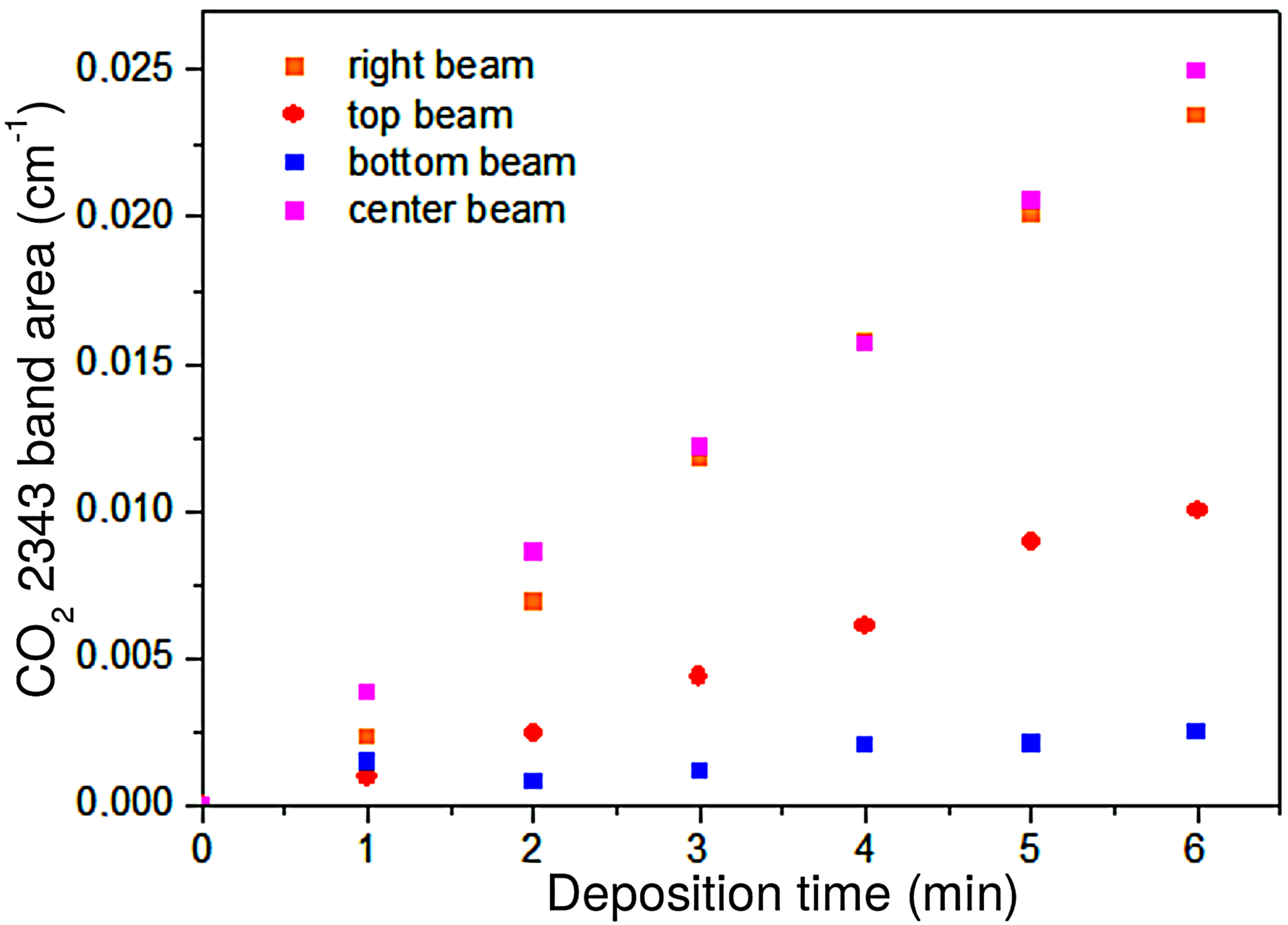}}}% Here is how to import EPS art
	\caption{\label{fig:beams2} Infrared band area as a function of CO$_{2}$ deposition time at a source feed rate of 0.2 sccm measured for the four beam lines.}
\end{figure}

In Figure~\ref{fig:beams2} we present the CO$_{2}$ IR band area as a function of deposition time obtained with the four beam lines. Each source was fed using a CO$ _{2}$ flux of 0.2~sccm. By comparing the QMS data shown in Figure~\ref{fig:beams1} and the IR data in Figure~\ref{fig:beams2} we can see that they are consistent. This indicates that the region of the cold target illuminated by IR light is exposed to the four beams, and the difference in IR band area only depends on flux intensity. The amount of deposited molecules grows linearly with time, the right and central beams are the most intense and their fluxes are comparable. The TB appears to have about one third of the flux intensity with respect to CB and RB, while the bottom beam has a very low flux, which is almost certainly due to a poor alignment. For this reason, at the time writing this paper, the bottom beam is so weak that cannot be used for experimental purposes, in that depositions would imply too long exposure times not compatible with a typical experimental sequence. From the integrated band areas of Figure~\ref{fig:beams2} at 5~min exposure, the relative intensities of CB, RB, TB, and BB are 1:1:0.33:0.1, respectively.

\begin{figure}[b] 
	%[scale=0.5] [width=0.5\hsize] 
	\includegraphics[scale=0.38]{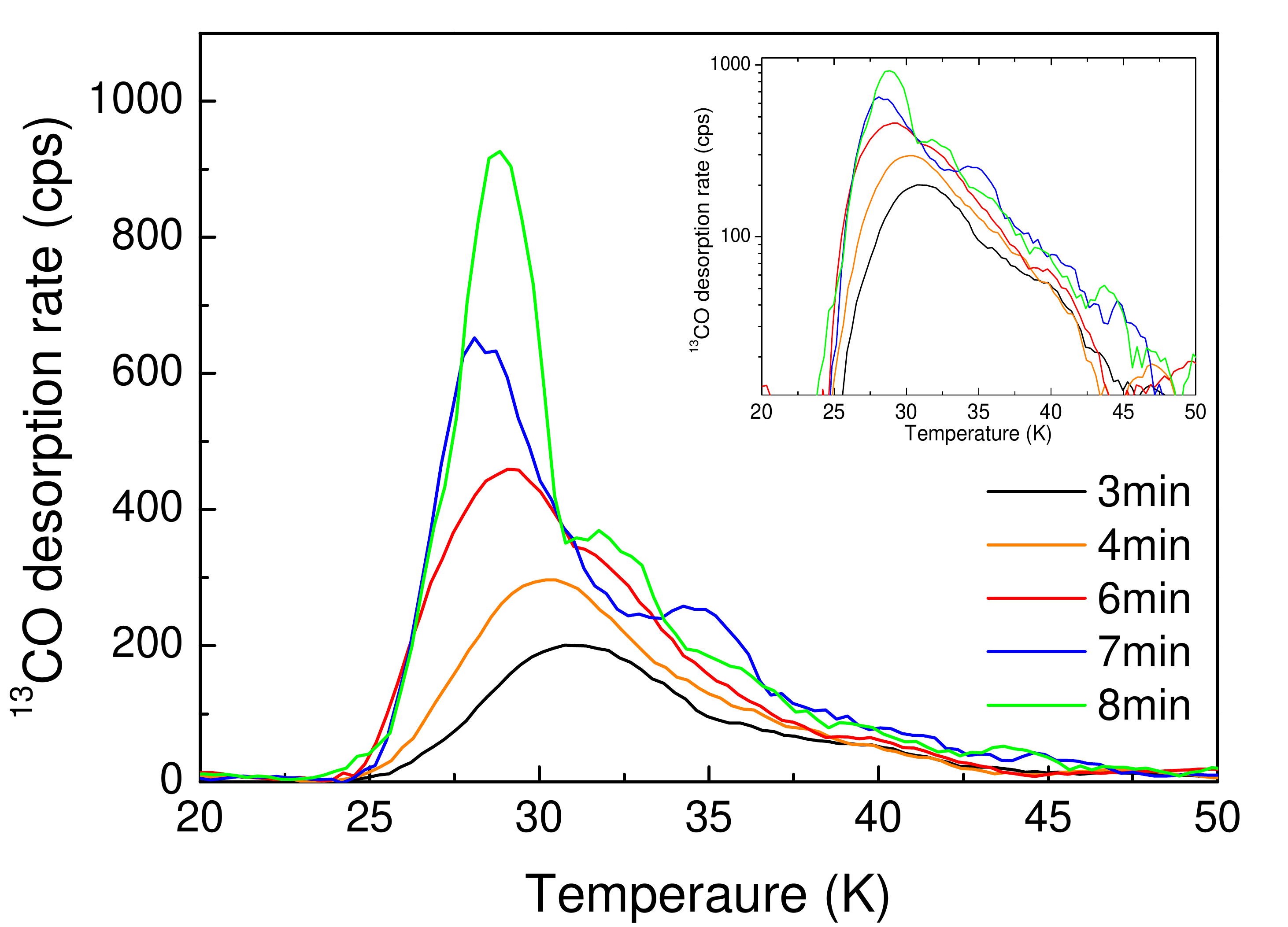}
	\caption{$^{13}$CO TPD series as function of surface temperature used for calibration of the central beam. \textit{Inset:} Logarithm of the desorption rate \textit{vs} temperature showing the exponential behavior of TPD traces corresponding to doses $>$1~ML. \label{fig:13COtpd}
	}
\end{figure}

Given a fixed source pressure, absolute beam flux calibration of the molecular beams can be performed by a series of TPD experiments using a fully wetting molecule, such as N$_{2}$, CO, or O$_{2}$, with sticking probability at low temperatures close to unity. Fully wetting species have molecule--substrate bonds stronger than molecule--molecule interactions. Hence, these species exhibit only (sub)monolayer desorption characteristics, and have no multilayer component for TPD spectra corresponding to 1~ML or less. \\ 
TPD mass spectra \textit{vs} surface temperature used for the flux calibration of the central beam are illustrated in Figure~\ref{fig:13COtpd}. The surface is cooled to 10~K and a timed deposition of $^{13}$CO is performed. The surface is heated and the number of molecules measured with the QMS as they desorb from the surface. The experiment is repeated and the dose is increased until the leading edge keeps shifting to the left, meaning that the surface is being filled with least binding sites being occupied last. From the $^{13}$CO TPD spectra, we notice that the desorption profile of CO stops shifting to the left after a dose of six minutes. From the inset in Fig.~\ref{fig:13COtpd}, showing the logarithm of the desorption rate vs surface temperature, it is clear that TPD profiles for doses $>6$~min increase exponentially from the same multilayer temperature (25~K), and a multilayer peak begins to grow on top of the full monolayer trace corresponding to 6 minutes. Therefore, $\sim$1~ML CO is deposited in 6~min, that is 360 seconds, which gives a CB flux: 

$\phi_{CB}=10^{15} \, \mathrm{molecules \, cm}^{-2} / 360 \, s \simeq  3 \times 10^{12} \, \mathrm{molecules \, cm}^{-2} \, s^{-1}$.\\
Because it mesures the amount of species deposited on the target, this method has the advantage of measuring the effective flux on the surface, rather than simply the flux of the beam.
From the values of relatives fluxes we can thus estimate the flux of the other beams:\\
$\phi_{RB} \simeq 3 \times 10^{12}\, \mathrm{molecules \, cm}^{-2} \, s^{-1}$,\\
$\phi_{TB} \simeq  1 \times 10^{12} \, \mathrm{molecules \, cm}^{-2} \, s^{-1}$,\\
$\phi_{BB} \approx 3 \times 10^{11} \, \mathrm{molecules \, cm}^{-2} \, s^{-1}$,\\
with a conservative error of 20\% on $\phi_{CB}$, $\phi_{RB}$, and $\phi_{TB}$, due to the uncertainty associated to the TPD spectra and the integrated infrared band areas.

%***

\subsubsection{\label{subsubsec:rairsconstant}Infrared absorption bands and column densities}

In general, for thin layers $\le 20$~ML, the column density $N$ (molecules cm$^{-2}$ ) of deposited or newly formed species can be calculated via a modified Lambert-Beer equation~\citep{bennett04,TEOLIS07}:

\begin{equation}
N = \Lambda \, \frac{\int_{\lambda_1}^{\lambda_2} A(\lambda) \mathrm{d}\lambda}{\mathrm{S}} \label{LambertBeer}
\end{equation}

where  $\int A(\lambda) \, \mathrm{d}\lambda$ is the integrated area of the infrared absorption feature (cm$^{-1}$), and S is the corresponding band strength. The constant $\Lambda$ is ${(\ln10 \,\,\, \mathrm{cos} \alpha)}/{2}$, 
where the division by 2 corrects for the ingoing and outgoing infrared beam, and $\alpha$ accounts for the angle between the normal of the surface mirror and the infrared beam. Therefore, the determination of the column density of the sample is apparatus dependent and literature values of band strengths ought to be used with care in RAIRS measurements. 
The setup dependent value of $\Lambda$  may be measured using two methods:

Method 1): computation of the theoretical value of $\Lambda$ and use of eqn.~\ref{LambertBeer} to derive $N$.

Method 2): a series of TPD experiments is performed to accurately determine the 1~ML dose of a given infrared active molecule. Subsequently, $N$ being known, the integrated absorbance of the infrared band can be measured and the value of $\Lambda$ for a given setup can be calculated from eqn.~\ref{LambertBeer}.
\\
As for the FT-RAIRS setup mounted on VENUS, with a grazing angle of 83$^\circ \pm 1^\circ$, the theoretical value is:

$\Lambda_{theo, VENUS} = 0.14 \pm 0.02$, \\
where the error represents the uncertainty in the angle. 

An experimental derivation of $\Lambda$ was done using the $^{13}$CO TPD spectra shown in Section~\ref{subsubsec:beamflux}.

For a coverage of 1~ML (10$^{15}$ molecules~cm$^{-2} \,   \pm$ 20\%), the integrated $^{13}$CO band at 2092~cm$^{-1}$ is $(0.006 \pm 0.001)$ cm$^{-1}$; S($^{13}$CO) is $1.3 \times 10^{-17}$ cm~molecules$^{-1}$~\cite{Gerakines1995}; $N$ must be corrected for the actual area of the spot on the sample. The circular spot has a $\sim$3 mm (0.3 cm) diameter, or 0.15~cm radius, which gives an area of $\pi r^2 = (0.07 \pm 0.01)$~cm$^{2}$. Hence, $N$ is 10$^{15}$ molecules cm$^{-2} \times 0.07 = 7 \times 10^{13}$~molecules~cm$^{-2}$. Therefore, by rearranging eqn.~(\ref{LambertBeer}), we obtain:

$
\Lambda_{exp, VENUS} = \frac{N \cdot \mathrm{S}}{\int A(\lambda) \mathrm{d}\lambda} = \frac{7 \times 10^{13} \mathrm{molecules} \, \mathrm{cm}^{-2} \, \cdot \, 1.3 \times 10^{-17} \mathrm{cm \, molecules}^{-1}}{0.006 \, \mathrm{cm}^{-1} } = 0.15 \pm 0.07. 
$

We can notice that the two calculated values of $\Lambda$ are consistent, although the theoretical value $\Lambda_{theo, VENUS}$ is affected by a smaller error. Hence, this value seems to be the most reliable for the quantitative measurements of column densities.

In the end, taking into account the value $\Lambda_{theo, VENUS}$ and the scaling factor for the beam size, the Lambert-Beer equation corrected for the calculation of the column density of species X with VENUS is:
\begin{equation}
N_{X} = \frac{0.14}{0.07} \, \frac{\int A(\lambda) \mathrm{d}\lambda}{\mathrm{S_{X}}} = 2 \, \frac{\int A(\lambda) \mathrm{d}\lambda}{\mathrm{S_{X}}}. \label{LambertBeerMOD}
\end{equation}

Considering all sources of error, a 50\% uncertainty on calculated column densities is likely to be conservative enough.
%***

\section{Experimental Performance} \label{sec:performance}

\subsection{Simple reactive systems as probes of beam spot overlap: the NH$_{3}$ + H$_{2}$CO reaction}

In this section we present the first experimental results obtained with the new multi-source apparatus called VENUS. The first representative results concern the thermal reactivity of H$_{2}$CO with NH$_{3}$, which had been previously investigated experimentally and theoretically~\cite{Schutte1993Icar,Bossa2009ApJ,Rimola2010PCCP}. This  reaction leads to the  formation of aminomethanol, NH$_{2}$CH$_{2}$OH, which occurs via a low barrier nucleophilic addition\footnote{The proposed general mechanism is a nucleophilic attack of the C atom of H$_{2}$CO by the lone pair of the N atom of NH$_{3}$, followed by a proton transfer from the N atom to the O atom~\cite{Courmier2005}.} between ammonia and formaldehyde:

\begin{equation}
\mathrm{NH}_{3} + \mathrm{H}_{2}\mathrm{CO} \longrightarrow \mathrm{NH}_{2}\mathrm{CH}_{2}\mathrm{OH}.
\label{react:aminomethanol}
\end{equation} 
In this first experiment we show how we can use the simple chemical system above \mbox{(A+B $\to$ AB)} to quantitatively assess the actual overlap of the beams on the target. NH$_{3}$ was deposited on the surface through the top beam. Formaldehyde is solid at ambient temperature so gaseous H$_{2}$CO was obtained by gently warming (at $\simeq 62^{\circ}$C) paraformaldehyde powder kept in a glass vial~\cite{Minissale2015}. The flux of H$_{2}$CO$_\mathrm{(gas)}$ is adjusted with a precision leak valve and injected into the central beam to be deposited onto the gold surface. Formaldehyde is characterized by two strong bands located at 1730 and 1500~cm$^{-1}$ assigned to the C=O stretching mode and the CH bending mode, respectively.

For each molecule, the characteristic bands are integrated, and then inserted into equation~(\ref{LambertBeerMOD}) with their corresponding band strengths to estimate their column densities.
For NH$_{3}$, we use the value of $1.1 \times 10^{-17}$ cm molecule$^{-1}$ for the band strength of the stretching mode at 3380~cm$^{-1}$~\cite{dHendecourt1986AAS}. For H$_{2}$CO we use the CO stretching mode band at 1730~cm$^{-1}$, which has a band strength of $9.6 \times 10^{-18}$ cm~molecule$^{-1}$~\cite{Schutte1993Icar}.

\begin{figure}[b] 
	%[scale=0.5] [width=0.5\hsize] 
	\includegraphics[scale=0.35]{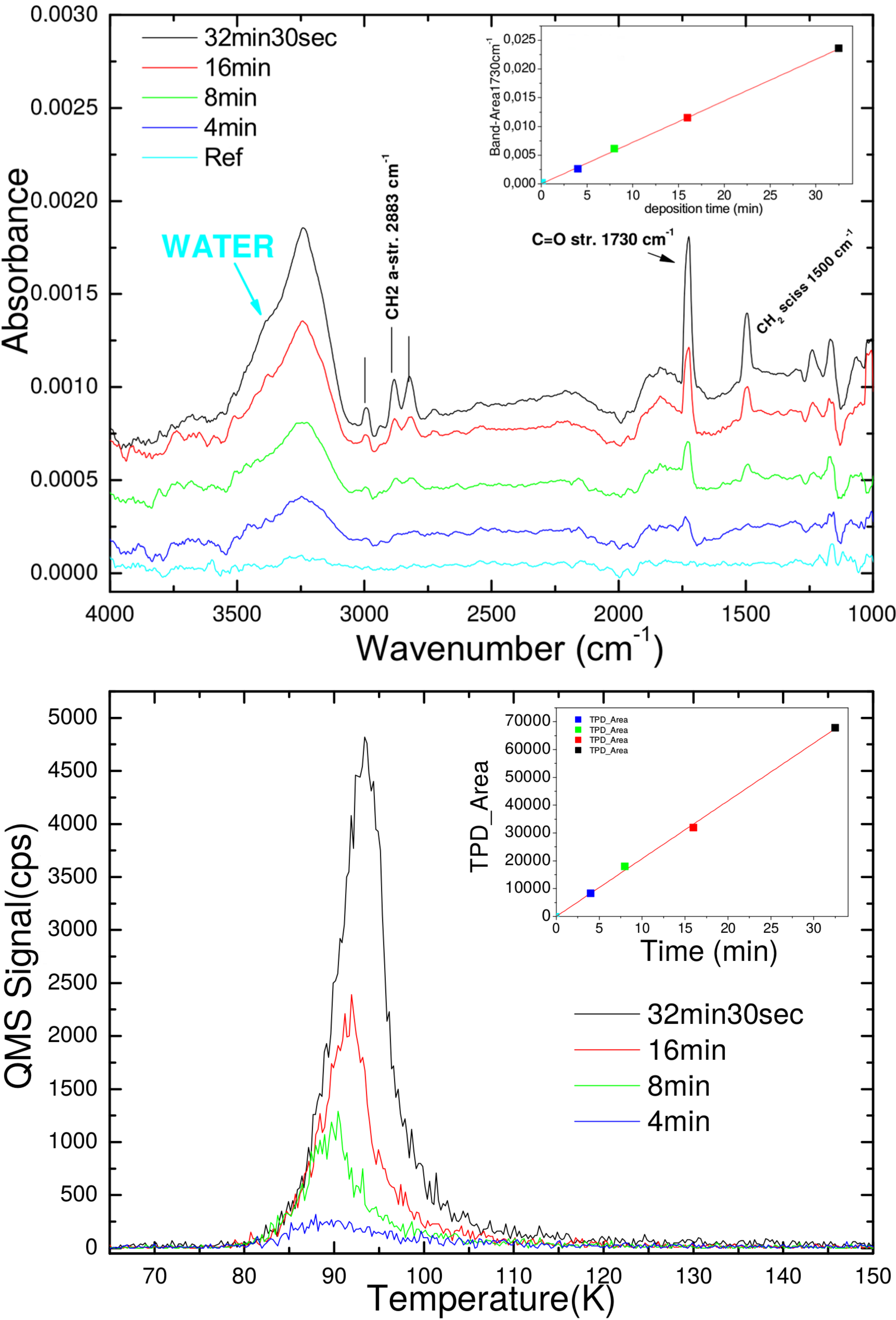}
	\caption{FT-IR spectra (top) of various timed depositions of formaldehyde at 65~K on the gold-plated target, and the subsequent TPD traces (bottom) recorded during the linear warmup at 12~K~min$^{-1}$. \textit{Insets}: H$_{2}$CO 1730~cm$^{-1}$ band area and the H$_{2}$CO TPD peak area, respectively, as a function of deposition time. \label{fig:h2co_ir_tpd}
	}
\end{figure}

\begin{figure}[b] 
	%[scale=0.5] [width=0.5\hsize] 
	\includegraphics[scale=0.38]{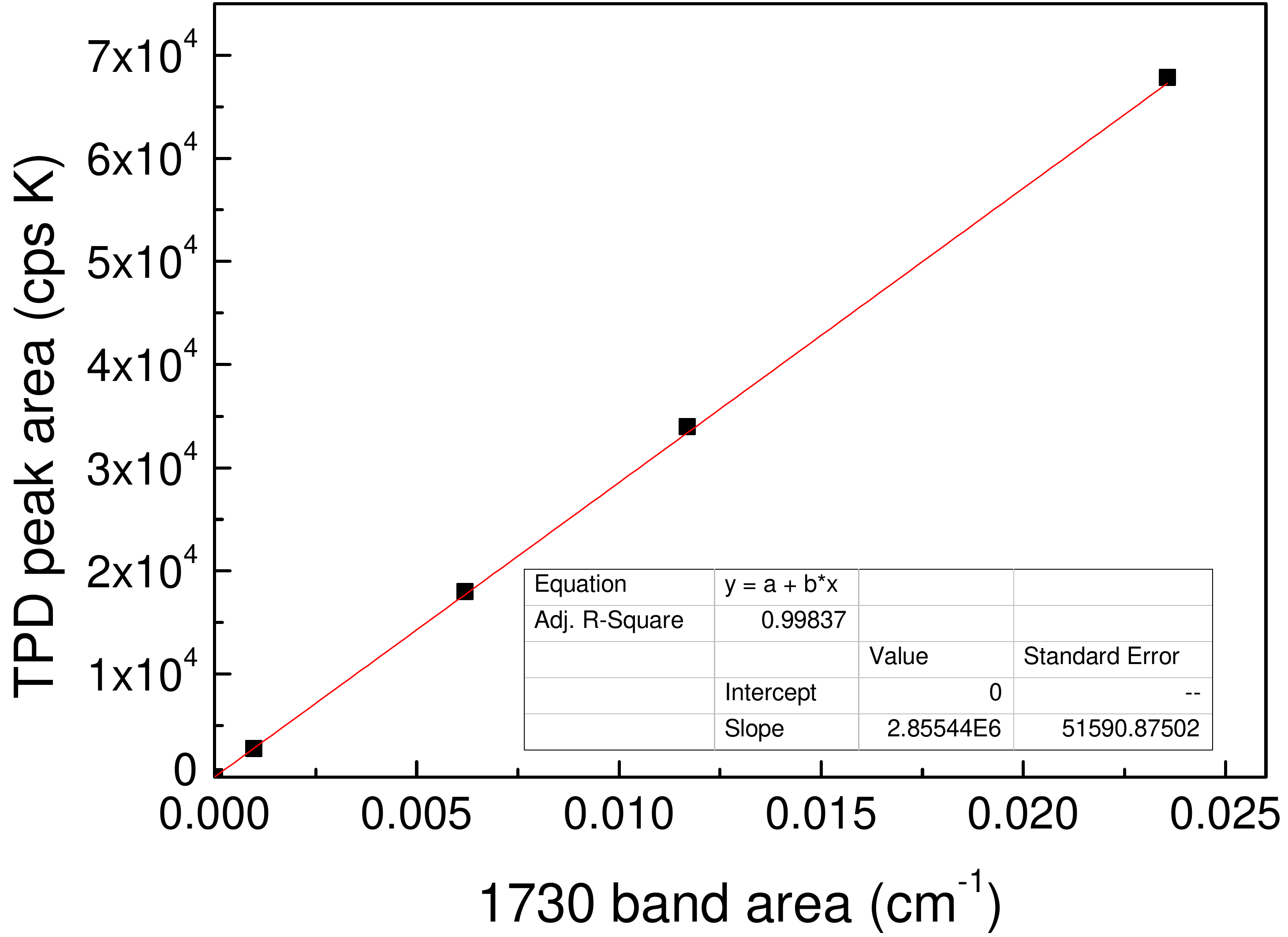}
	\caption{Correlation between TPD peak areas and 1730~cm$^{-1}$ band areas of H$_{2}$CO. Each data point corresponds to a given deposition time as per experiments presented in Fig.~\ref{fig:h2co_ir_tpd}. \textit{Inset}: details of the linear fit plotted in red.  \label{fig:QMS-RAIRS}
	}
\end{figure}

In Figure~\ref{fig:h2co_ir_tpd}, we show a series of FT-IR spectra (top image) of various timed depositions of formaldehyde at 65~K on the gold-plated target, and the subsequent TPD traces (bottom image) recorded during the linear warmup at 12~K~min$^{-1}$. Both images contain an inset showing the H$_{2}$CO 1730~cm$^{-1}$ band area and the H$_{2}$CO TPD peak area, respectively, as a function of deposition time. We can see that both trends are linear, which indicates that \textbf{a)}~the amount of molecules actually deposited is proportional to the dosing time and \textbf{b)}~qualitatively and quantitatively the two analytical tools implemented on VENUS are consistent (see Fig.\ref{fig:QMS-RAIRS}).

\begin{figure}[b] 
	%[scale=0.5] [width=0.5\hsize] 
	\includegraphics[scale=0.4]{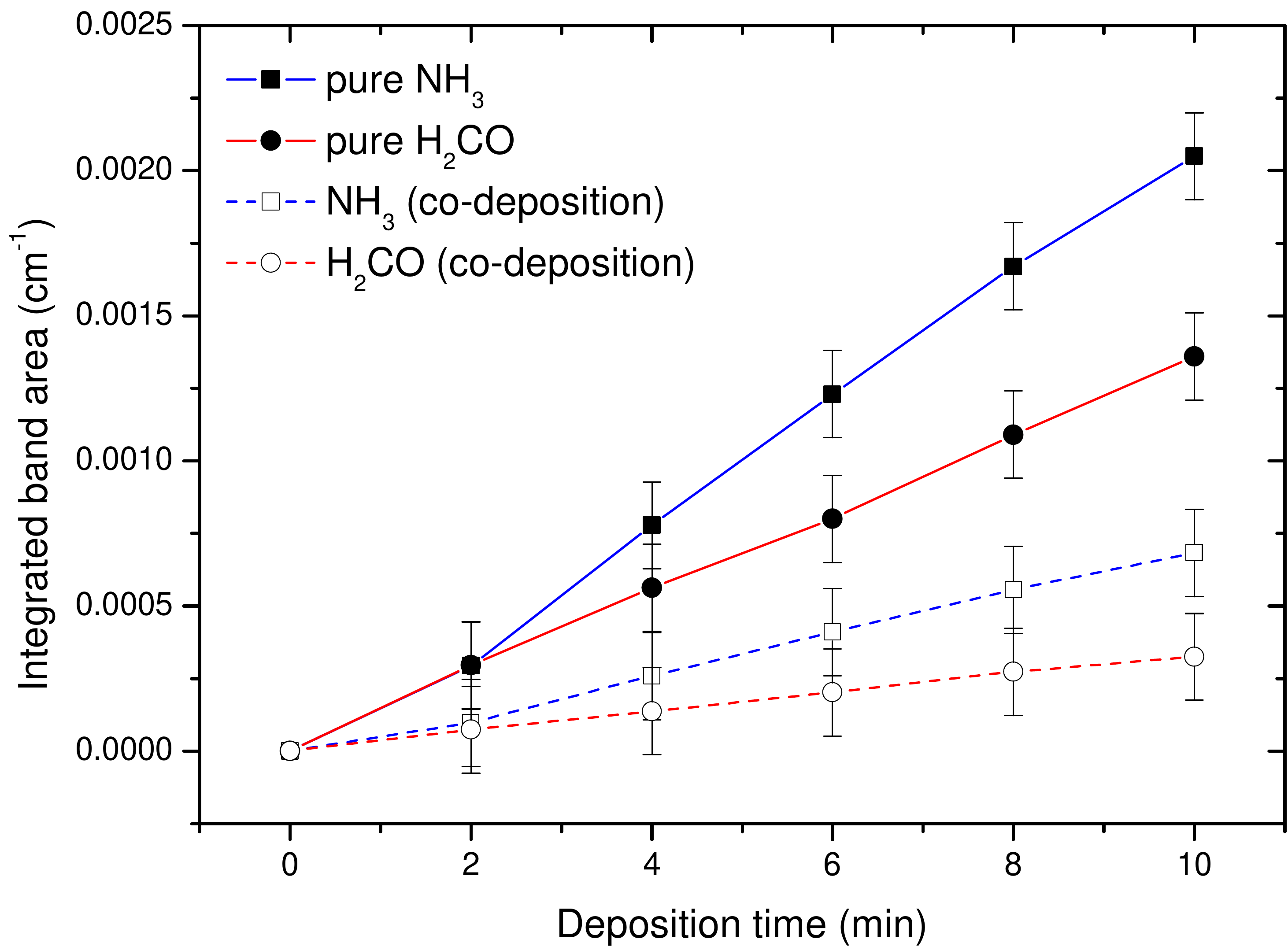}
	\caption{Integrated infrared band area of NH$_{3}$ and H$_{2}$CO in pure ices experiments (filled squares and circles), and in co-depositions of NH$_{3}$ + H$_{2}$CO (empty square and circles) at a surface temperature of 65~K, as a function of deposition time.
    \label{fig:IRNH3-H2C0}
	}
\end{figure}

Figure~\ref{fig:IRNH3-H2C0} shows the integrated IR band area of NH$_{3}$ or H$_{2}$CO pure ices for various deposition times, and of various doses of co-deposited NH$_{3}$ and H$_{2}$CO at a surface temperature of 65~K. We can use the integrated band area of the pure ices to estimate the deposited column densities via TB (NH$_{3}$) and CB (H$_{2}$CO). For a dose of 10~min, A$_{NH_{3}}$ is 0.00205~cm$^{-1}$ and A$_{H_{2}CO}$ is 0.00136 cm$^{-1}$ (see Table~\ref{tab:IRNH3-H2C0}). Putting these values in equation~(\ref{LambertBeerMOD}), we obtain N$_{NH_{3}}$ = $3.7 \times 10^{14}$ mol~cm$^{-2}$ and N$_{H_{2}CO}$ = $2.8 \times 10^{14}$ mol~cm$^{-2}$, which gives a relative flux $\Phi_{NH_{3}}$\,:\,$\Phi_{H_{2}CO}$\,$\simeq$\,1\,:\,0.76, that is co-depositions will result in NH$_{3}$\,:\,H$_{2}$CO mixtures with an excess of ammonia.

\begin{table*}
	\caption{\label{tab:IRNH3-H2C0}Integrated IR band areas of NH$_{3}$ and H$_{2}$CO obtained after timed depositions of either pure species or codepositions at a surface temperature of 65~K. The error on all band areas is $\pm$1.5e-4~cm$^{-1}$.}
	\begin{ruledtabular}
		\begin{tabular}{ccccc}
			&\multicolumn{2}{c}{Pure species}&\multicolumn{2}{c}{NH$_{3}$\,+\,H$_{2}$CO experiment}\\
			Dose (min)&A$_{NH_{3}}$\,(cm$^{-1}$)&A$_{H_{2}CO}$\,(cm$^{-1}$)&A$_{NH_{3}}$\,(cm$^{-1}$)&A$_{H_{2}CO}$\,(cm$^{-1}$)\\ \hline
			2&2.95e-4&2.96e-4&9.83e-5&7.40e-5 \\
			4&7.78e-4&5.63e-4&2.59e-4&1.38e-4\\
			6&1.23e-3&8.00e-4&4.10e-4&2.05e-4\\
			8&1.67e-3&1.09e-3&5.56e-4&2.73e-4\\
			10&2.05e-3&1.36e-3&6.83e4&3.25e-4\\
		\end{tabular}
	\end{ruledtabular}
\end{table*}

From Fig.~\ref{fig:IRNH3-H2C0}, it is also clear that smaller amounts of NH$_{3}$ or H$_{2}$CO are measured on the surface when the two species are co-deposited, which indicates that a reaction takes place and both ammonia and formaldehyde are consumed.
Schutte, Allamandola, and Sandford (1993)~\cite{Schutte1993Icar} have shown that NH$_{2}$CH$_{2}$OH is the dominant product when NH$_{3}$ is in excess with respect to H$_{2}$CO, and we will use this argument to make the following hypotheses: 

\textit{i)} working with an excess of NH$_{3}$ impedes formaldehyde polymerisation;

\begin{figure}[b] 
	%[scale=0.5] [width=0.5\hsize] 
	\includegraphics[scale=0.4]{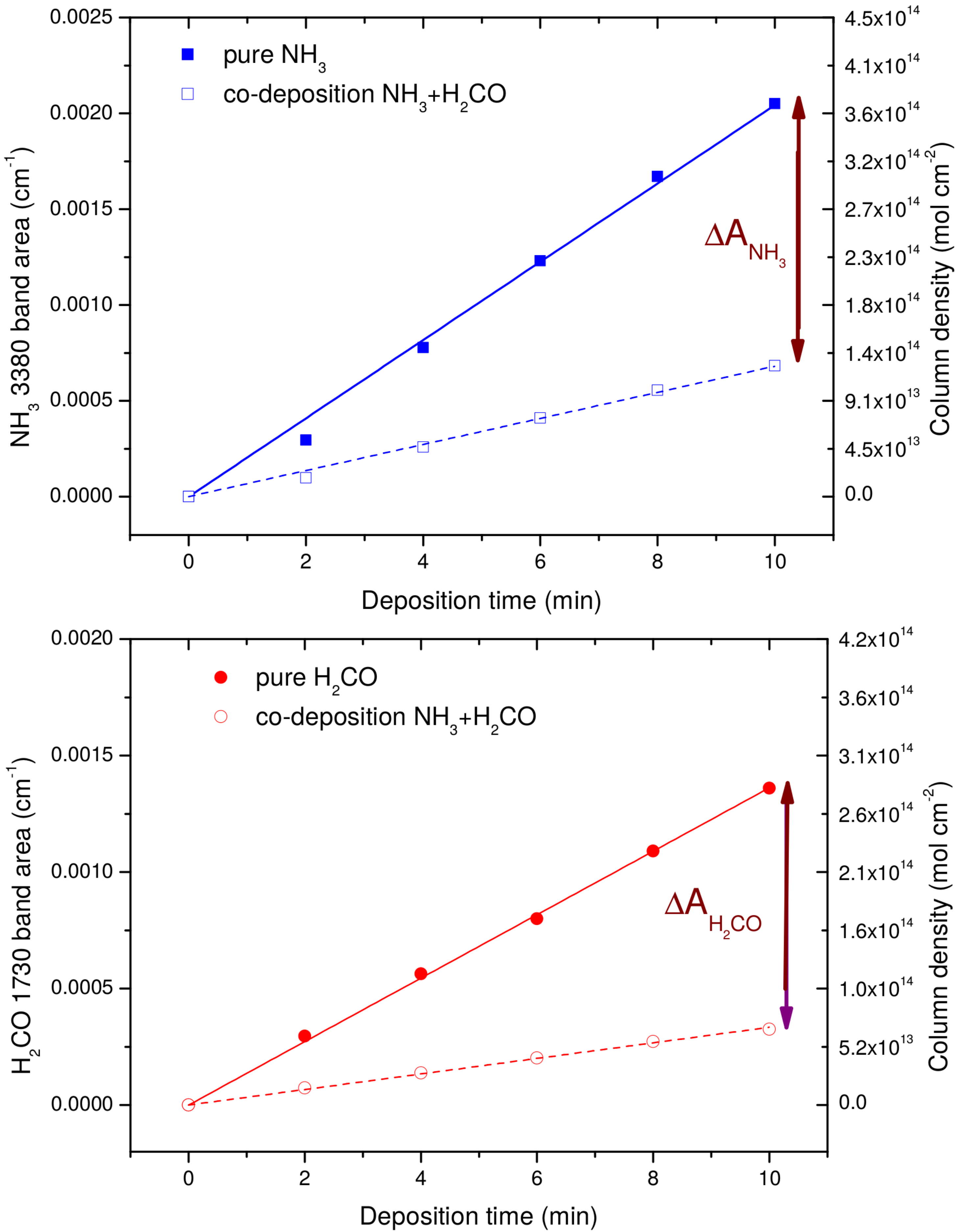}
	\caption{\textit{Top panel:} integrated infrared band area of NH$_{3}$ for deposition of NH$_{3}$ only (filled squares) and co-deposition NH$_{3}$ + H$_{2}$CO (empty squares), as a function of deposition time. \textit{Lower panel:} integrated infrared band area of H$_{2}$CO for deposition of H$_{2}$CO only (filles circles) and co-deposition NH$_{3}$ + H$_{2}$CO (empty circles), as a function of deposition time. Solid and dashed lines represent linear fits of the data. \label{fig:H2COcomp}
	}
\end{figure}

\textit{ii)} formaldehyde is completely consumed if the two beams overlap.\\
A quantitative estimate of the amount of NH$_{3}$ and H$_{2}$CO consumed during the co-deposition experiment, with respect to the pure ice case, is illustrated in Figure~\ref{fig:H2COcomp}. 

$\Delta$A is the difference between the integrated band area of the pure ice and that measured after the NH$_{3}$ + H$_{2}$CO experiment, for a deposition time of 10 minutes. From the values of $\Delta$A$_{NH_{3}}$ and $\Delta$A$_{H_{2}CO}$, using eqn.~(\ref{LambertBeerMOD}), it is straightforward to calculate the  amounts of adsobates consumed in the reaction, $\Delta$N(NH$_{3}$) and $\Delta$N(H$_{2}$CO), respectively:

\begin{equation}
\Delta N(\mathrm{NH}_{3}) = 2 \, \frac{\Delta  \mathrm{A}(\mathrm{NH}_{3})_{10'}}{S_{\mathrm{NH}{3}}} = (2.5 \pm 0.3) \times 10^{-14} \, \mathrm{molecules \,\, cm}^{-2},
\end{equation}
	
and

\begin{equation}
\Delta N(\mathrm{H}_{2}\mathrm{CO}) = 2 \, \frac{\Delta  \mathrm{A}(\mathrm{H}_{2}\mathrm{CO})_{10'}}{S_{\mathrm{H}{2}CO}} = (2.2 \pm 0.5) \times 10^{-14} \, \mathrm{molecules \,\, cm}^{-2}. 
\end{equation}

Since $\Delta$N(NH$_{3}$) $\simeq$ $\Delta$N(H$_{2}$CO) within experimental errors, ammonia and formaldehyde are consumed at the same rate, which confirms that reaction~(\ref{react:aminomethanol}) is at play. Hence, in experiments with an excess of NH$_{3}$ where H$_{2}$CO would react and be completely consumed, we can derive the maximum percentage of overlap of the two beams by calculating the ratio between $\Delta$A$_{H_{2}CO}$ and A$_{H_{2}CO}$. For the 10-min experiment, we found 

\begin{equation}
\frac{\Delta\mathrm{A}(\mathrm{H}_{2}\mathrm{CO})_{10'}} {\mathrm{A}(\mathrm{H}_{2}\mathrm{CO})_{10'}} 
= \frac{2.2 \times 10^{-14}}{2.8 \times 10^{-14}} = 0.78 \simeq 78\%. 
\end{equation}

\begin{figure}[b] 
	%[scale=0.5] [width=0.5\hsize] 
	\includegraphics[scale=0.37]{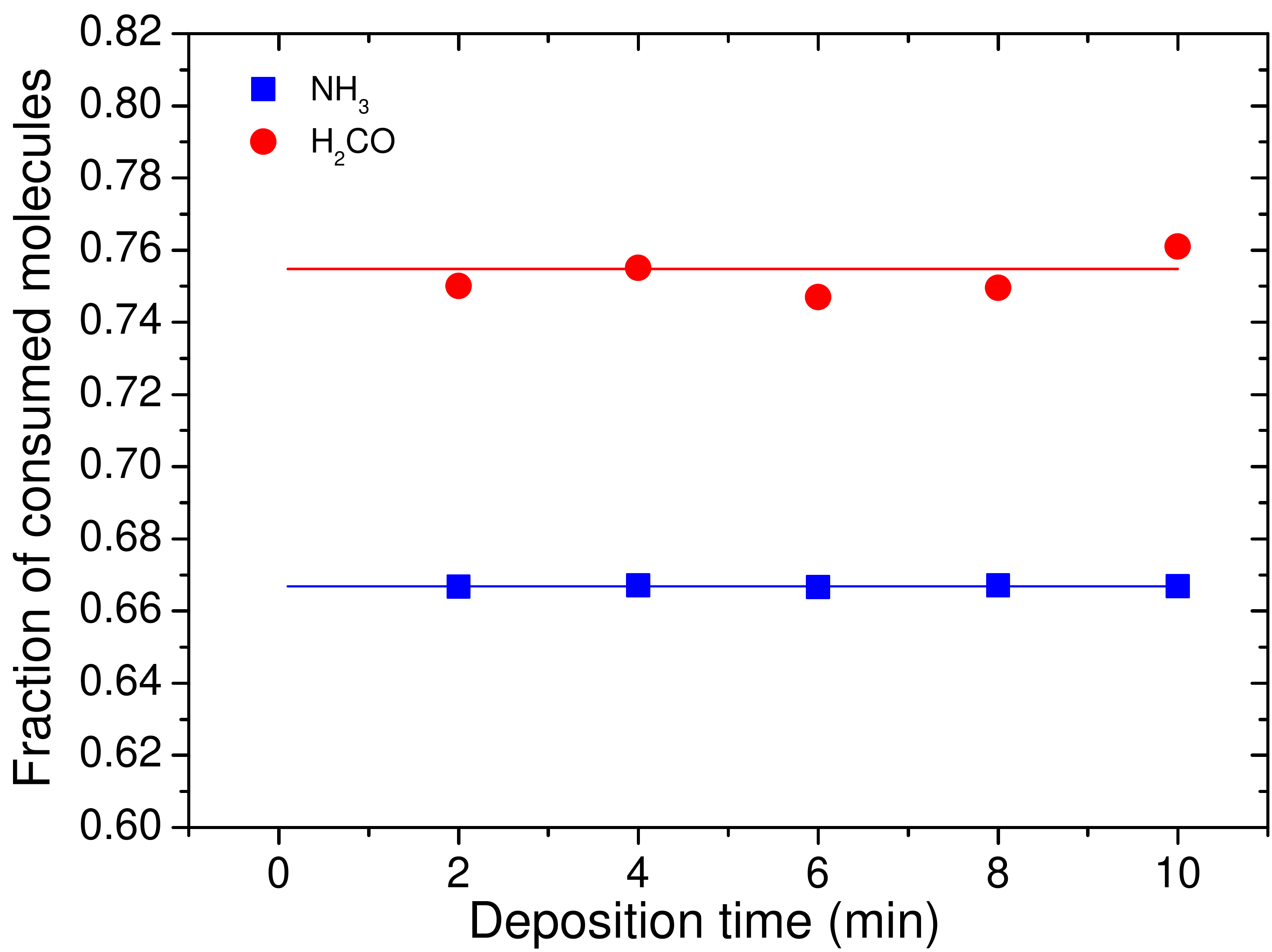}
	\caption{Fraction of consumed adsorbates ($\Delta$A$_{NH_{3}}$/A$_{NH_{3}}$, blue squares; $\Delta$A$_{H_{3}CO}$/A$_{H_{3}CO}$, red circles) as a function of co-deposition time in the NH$_{3}$ + H$_{2}$CO experiment at 65~K. $\Delta$A$_{H_{3}CO}$/A$_{H_{3}CO} \approx 75\%$ also represents the degree of overlap of the two beam spots. Solid lines are linear fits of the data. \label{fig:overlap}
	}
\end{figure}

Extending this analysis to all the experiments done at different doses, we can plot a more general trend and thus reduce the error bar on the estimate of the overlap of beam spots TB and CB. This trend is presented in Figure~\ref{fig:overlap}. We can observe that the fraction of deposited H$_{2}$CO that reacts on the surface, hence the percentage of overlap of the two beams, is $\approx$~75~\%. On the other hand, as one would expect in the case of experiments carried out in excess of ammonia, the fraction of NH$_{3}$ molecules involved in the reaction is smaller, that is $\approx$~67~\%. By scaling the two fractions of consumed molecules of NH$_{3}$ and H$_{2}$CO with the corresponding band strength, it is easy to check that $\Delta$N(NH$_{3}$)$_\mathrm{react}$/$\Delta$N(H$_{2}$CO)$_\mathrm{react} \sim 1$.   

%\newpage

\subsection{Chemical layer: the NO + H reaction system}

The first original scientific results obtained using VENUS concern the NO~+~H reaction, with the objective to investigate whether H atoms can penetrate into NO ices beyond the first outer monolayer. 
In surface reactions and solid-state chemistry, nitric oxide, NO, is thought to be the main precursor species of the NO-bonded molecules~\cite{Halfen_2001} and, together with NH$_{3}$, the precursor molecule of nitrogen-containing organics~\cite{Tsegaw2017,Blagojevic2003MNRAS}. Recently, millimeter-band observations of a molecular cloud uncovered the presence of an essential RNA precursor in space, hydroxylamine (NH$_{2}$OH), whose formation mechanism is predicted to be by chemical models and by laboratory experiments the hydrogenation of NO on dust grain surfaces~\cite{Rivilla2020}.

\begin{figure}[b] 
	%[scale=0.5] [width=0.5\hsize] 
	\includegraphics[scale=0.49]{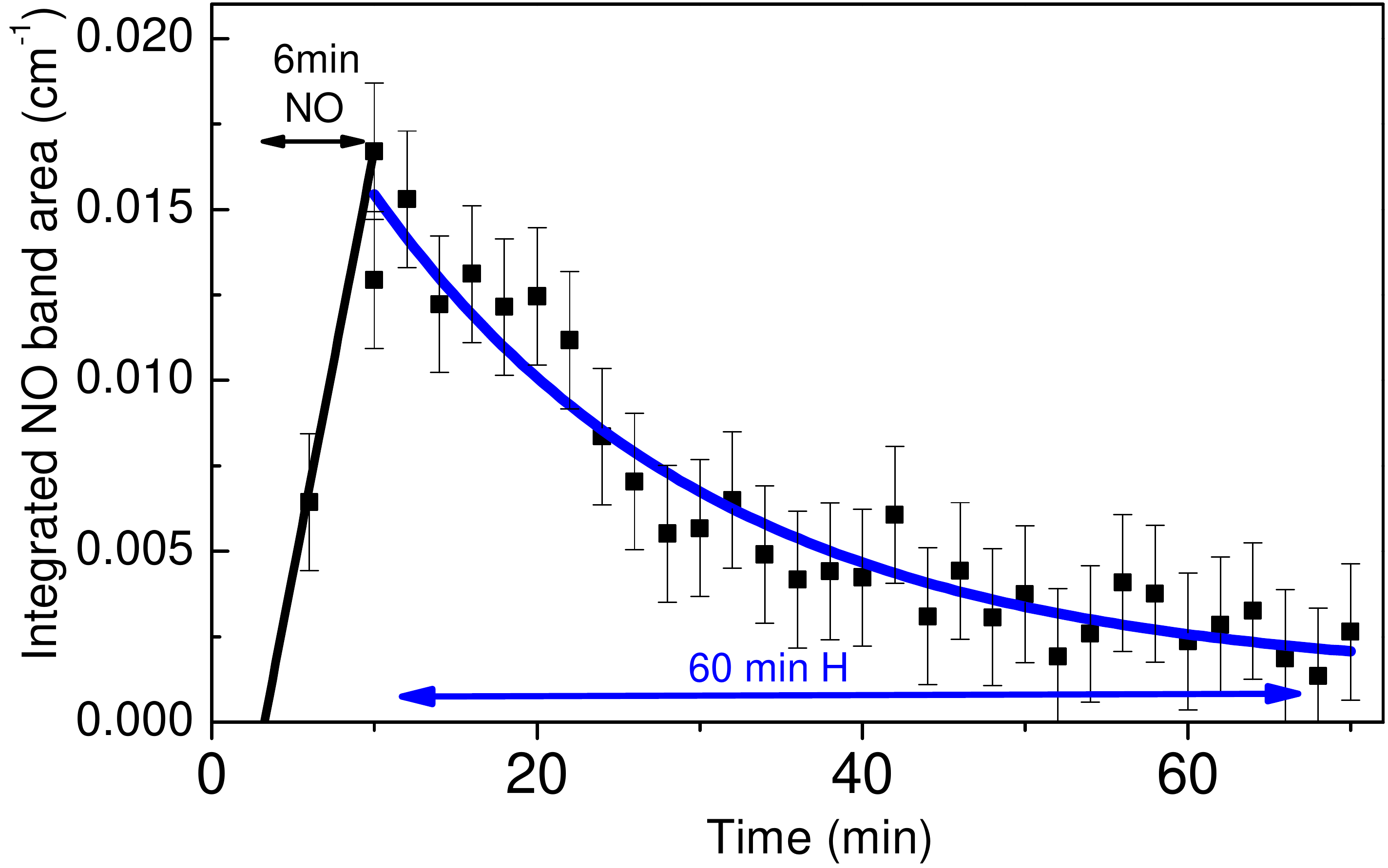}
	\caption{Integrated IR band area of NO (black squares) measured during the deposition phase (black line, 6~min$\simeq$1~ML), and during the H atom exposure time (60 min) fitted with exponential function (blue trace). \label{fig:1NO_Hreactivity}
	}
\end{figure}

Figure~\ref{fig:1NO_Hreactivity} shows the integrated area of the NO dimer band at 1775~cm$^{-1}$ as a function of deposition time. The black solid line represents the band area during the deposition of 1~ML (6 min dose) of NO at 10~K on the bare gold substrate, while the subsequent exponential decrease starting at t = 10~min shows the evolution of the NO band area during H bombardment. This experiment confirms that the NO molecule reacts rapidly with H atoms~\cite{Congiu2012ApJ,Congiu2012JChPh} and also shows that the FT-IR spectrometer can probe very thin layers of adsorbates, with a sensitivity of at least 0.2  monolayers. 

\begin{figure}[t] 
	%[scale=0.5] [width=0.5\hsize] 
	\includegraphics[scale=0.49]{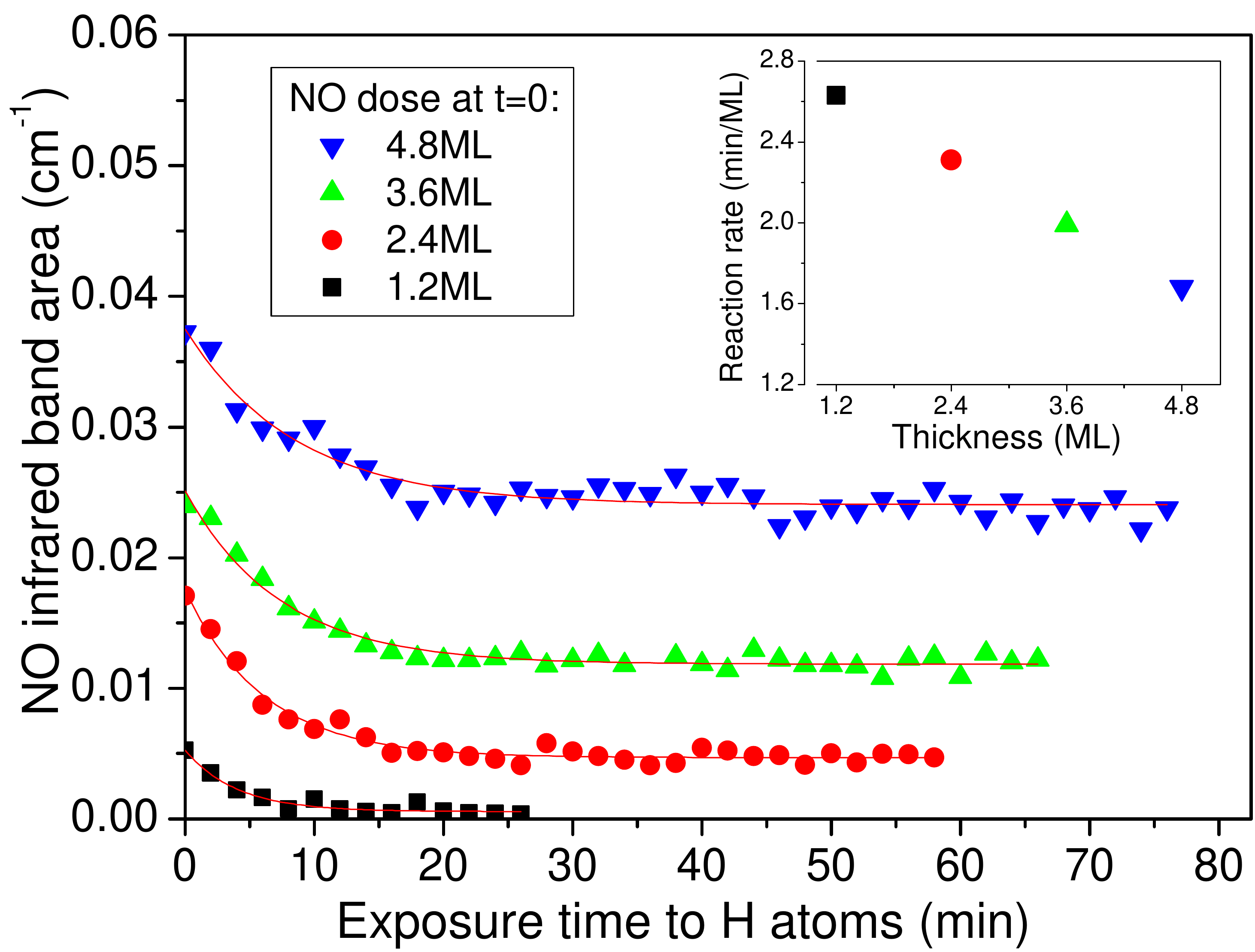}
	\caption{Evolution of the infrared band area of NO with H atom bombardment for various initial doses of NO. \textit{Inset}: reaction rate indicating the consumption of NO as a function of thickness. \label{fig:2NO_Hreactivity}
	}
\end{figure}

In Figure~\ref{fig:2NO_Hreactivity} we present the consumption of a NO ice substrate of various initial thickness. A series of depositions (1.2, 2.4, 3.6, 4.8~ML) was performed and each NO ice sample was exposed to H atoms. The initial integrated band area decreases until a plateau is reached. For each of the four experiments, the decrease is well fitted with an exponential function, namely \mbox{Y = Y$_{0}$ + A$_{0}$\,exp(-t/$\tau$)},
whose corresponding reaction rates expressed in minutes/ML are reported in the inset of Fig.~\ref{fig:2NO_Hreactivity}. The correlation between the reactivity and thickness indicates that the thinner the ice the faster the consumption of NO molecules.

\begin{figure}[b] 
	\includegraphics[scale=0.49]{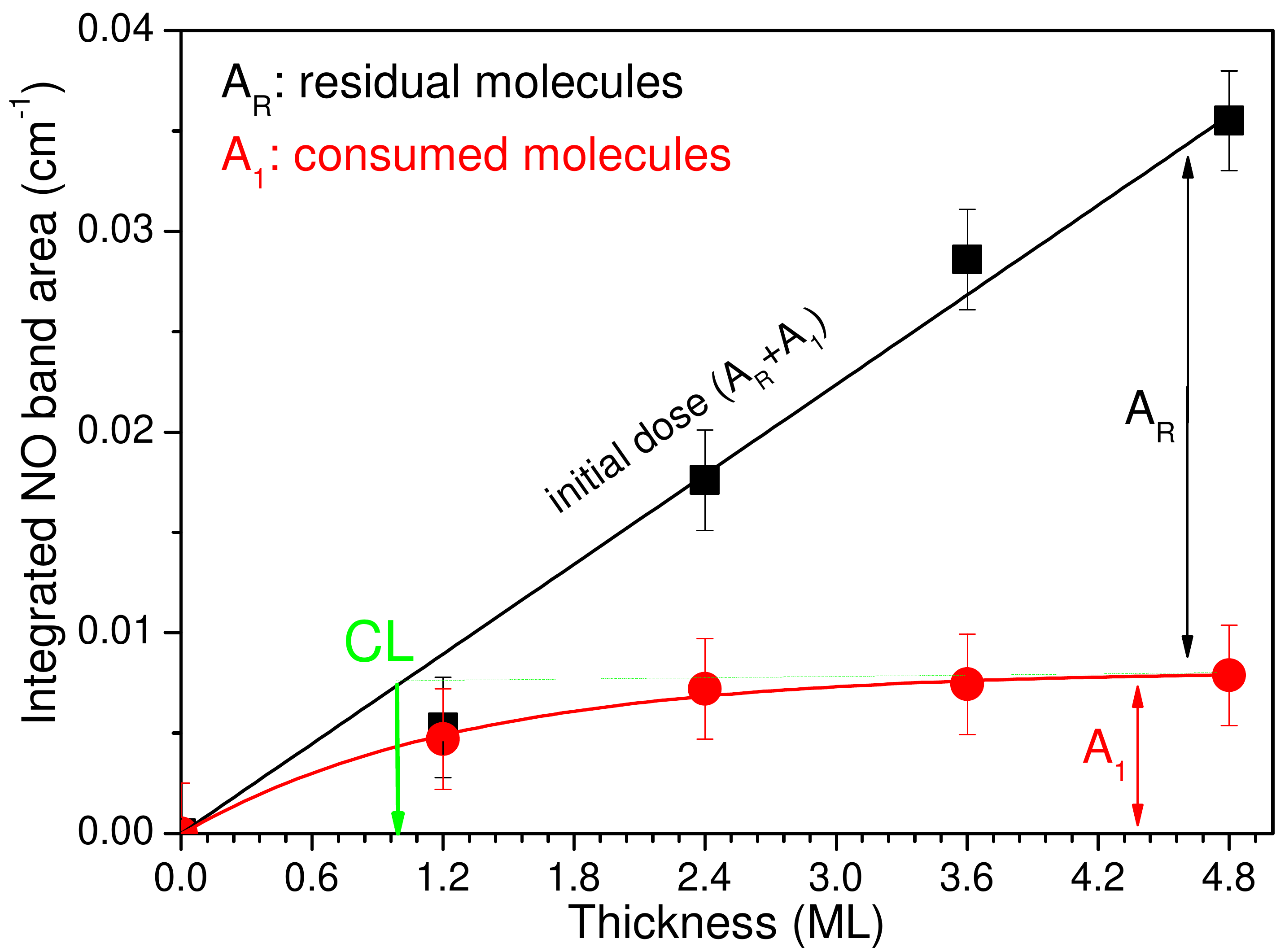}
	\caption{Comparison between deposited NO (A$_{0}$=A$_\mathrm{R}$~+~A$_1$, black squares) and the amount of NO consumed during H bombardment (A$_1$, red circles) as a function of ice thickness. The green horizontal line, representing the asymptotic limit of the exponential fit of A$_1$, intersects the black line at a thickness value of $\sim$1~ML coverage, which therefore corresponds to the chemical layer (CL).
	\label{fig:3NO_Hreactivity}
	}
	
\end{figure}

Also, by subtracting the area reached at t~=~$\infty$ (A$_\mathrm{R}$) from the initial band area, we can calculate the amount of NO that has reacted with H (A$_1$). This result is shown in Figure~\ref{fig:3NO_Hreactivity}. Black squares represent the typical infrared band area (A$_\mathrm{R}$~+~A$_1$) as a function of NO ice initial thickness. Red circles show the amount A$_1$, that is the fraction of NO molecules that have reacted with H atoms. We can see that A$_1$ never surpasses 1~ML thickness although it tends to it asymptotically, as shown by the horizontal green line that indicates the~0.008~cm$^{-1}$ value corresponding to $\sim$1 ML coverage.

These results clearly show that only the outer layer of a slab of NO ice reacts with H atoms coming from the gas phase, meaning that H atoms cannot penetrate beyond the first outer layer of NO. In other terms, the first physical molecular layer of ice coincides with the only accessible chemical layer that is consumed by the reaction NO~+~H to form NH$_{2}$OH, N$_{2}$O, H$_{2}$O~\cite{Thanh2019ECS}. At 10~K these new products are likely to form a barrier shielding the underlying layer from H atoms that cannot penetrate deeper. Moreover, the NO~+~H reaction proceeds faster in the case of thin layers of NO ice, which is likely to be the same conditions found on cosmic dust grains as corroborated by the recent detection of NH$_{2}$OH for the first time, toward a quiescent molecular cloud located in the Galactic Center~\cite{Rivilla2020}.
We obtained similar results with pure ices of formaldehyde exposed to H atoms.

This finding suggests that chemistry at the surface of dust grains only occurs in the first layer of a substrate of NO ice under our experimental conditions, although such results may be more general and apply to other chemical systems of astrophysical interest.

\newpage

\section{Astrophysical potential}\label{sec:potential}

VENUS, the apparatus described in this paper, will be used to study radical-radical reactions and investigate the reaction-diffusion processes that lead to the formation of complex organic molecules on icy grains, that is the formation of the building blocks of life in space. The new experiment has been constructed and successfully commissioned and calibrated. We have already published a Letter investigating the four-species system \{H+H$_2$CO+NO+H$_2$O\}. In Dulieu et al.~\cite{Dulieu2019MNRAS} we demonstrate how formamide (NH$_2$CHO), one of the key molecules of prebiotic chemistry~\cite{Saladino2012}, can be formed by hydrogenation of a mixture of H$_2$CO and NO, with water molecules playing a catalytic role, hence demonstrating the additional potential of an experiment built to study surface reactions using four deposition lines. 

VENUS will offer the possibility to direct simultaneously or sequentially atoms and molecular radicals onto an analog of cosmic dust. Its multi-beam configuration will allow us to deposit different ice substrates and investigate at once three-radical systems, while keeping a fully controlled experiment and ensuring maximum purity of the reactants. 

{Provided a second microwave cavity is installed,} N-bearing radicals such as NH and NH$_{2}$ can be formed \textit{in situ}, as well as OH and CH$_{3}$ groups. These radicals can be mixed and/or their reactivity and binding energy can be measured on previously deposited pure ices (i.e., H$_{2}$O, H$_{2}$CO, CO, NO) or ice mixtures. Also, by varying some selected experimental parameters (beam fluxes, surfaces temperature, and deposition times), it is possible to distinguish what are the active reactive pathways and estimate reaction rates from the analysis of the number and abundance of the products.

Furthermore, laboratory data on complex reaction systems, as long as they are fully controlled, will be complementary to astrochemical models and the increasingly high-resolution observations of interstellar ices in dark clouds and in circumstellar environments. In particular, with the imminent launch of the James Webb Space Telescope (JWST), a 6.5-meter orbiting infrared observatory scheduled for 2021~\cite{JWST2021}, astronomers will be able to look with greatly improved sensitivity inside dust clouds where stars and planetary systems form, that is environments where COMs are likely to be synthesized on icy grains. This will yield in the near future a wealth of data on the composition of interstellar ices and then it will be important to be able to produce appropriate experimental results, whose inclusion in astrochemical models will give us the most complete understanding possible on the formation of complex molecules in space.

\newpage

\begin{acknowledgments}
This work was supported by the Programme National “Physique et Chimie du Milieu Interstellaire” (PCMI) of CNRS/INSU with INC/INP co-funded by CEA and CNES, by the DIM ACAV+ a funding program of the Region Ile de France, and by the ANR SIRC project (GrantANR-SPV202448 2020-2024). T.N. gratefully acknowledges support from the LabEx MICHEM.
%\dots.
\end{acknowledgments}

\section*{Data availability}
The data that support the findings of this study are available from the corresponding author upon reasonable request.

%\nocite{*} %NOCITE* shows all refs within file.bib
%\bibliographystyle{plain}
\bibliography{VENUSpaper}% Produces the bibliography via BibTeX.

\end{document}